\documentclass[
 a4paper,
 superscriptaddress,
 notitlepage,
 amsmath,amssymb,
 aps,
 twocolumn,
 prl
 floatfix
]{revtex4-1}

\usepackage{tikz}
\usetikzlibrary{calc,arrows,decorations.pathreplacing,backgrounds}

\colorlet{mylinkcolor}{blue!66!black!80}

\usepackage{mathtools,amssymb,amsfonts}

\usepackage[colorlinks=true,linkcolor=mylinkcolor,citecolor=mylinkcolor,filecolor=cyan,urlcolor=mylinkcolor,breaklinks=true]{hyperref}
\usepackage{xcolor}
\usepackage[utf8]{inputenc}
\usepackage{booktabs}
\usepackage{bbm,mathrsfs,mathtools}
\usepackage{graphicx}

\newcommand{\e}[1]{\mathrm{e}^{#1}}

\newcommand{\bx}{\mathbf{x}}

\newcommand{\blue}[1]{\textcolor{black}{#1}}

\DeclareMathOperator{\arcsinh}{arcsinh}

\begin{document}
\title{Thermodynamically Consistent Phase-Field Theory Including Nearest-Neighbor Pair
  Correlations Explains Failure of Mean-Field Reasoning}

\author{Kristian Blom}
\affiliation{Mathematical bioPhysics group, Max Planck Institute for Multidisciplinary Sciences, G\"{o}ttingen 37077, Germany}

\author{Noah Ziethen}
\affiliation{Theory of Biological Fluids, Max Planck Institute for Dynamics and Self-Organization, G\"{o}ttingen 37077, Germany}

\author{David Zwicker}
\affiliation{Theory of Biological Fluids, Max Planck Institute for Dynamics and Self-Organization, G\"{o}ttingen 37077, Germany}

\author{Alja\v{z} Godec}
\email{agodec@mpinat.mpg.de}
\affiliation{Mathematical bioPhysics group, Max Planck Institute for Multidisciplinary Sciences, G\"{o}ttingen 37077, Germany}

\date{\today}

\begin{abstract}
Most of our current 
understanding of phase separation is
based on ideas that disregard
correlations. Here we 
illuminate unexpected effects of correlations on the structure
and thermodynamics of interfaces and in turn phase separation, which
are decisive in systems with strong interactions. Evaluating the
continuum limit of the
Ising model on the
Bethe-Guggenheim level, we derive a
Cahn-Hilliard free energy
that takes into
account
pair correlations. For a one-dimensional interface in a strip geometry these are shown to
give rise to an \emph{effective interface broadening} at
interaction strengths near and above the thermal energy,
which is verified in the Ising
model.\ Interface broadening is the result of an entropy-driven
  interface delocalization, which is not accounted for in the widely adopted
  mean field theory.\ Pair correlations \blue{are required for
    thermodynamic consistency} as they enforce a thermodynamically
  optimal \blue{local}
configuration of defects and profoundly affect
  nucleation and spinodal decomposition at strong coupling.
\end{abstract}
\maketitle
\section{Introduction}
Instigated by the seminal works of Cahn and
Hilliard \cite{cahn1958freeI, cahn1959freeII, cahn1959freeIII}, phase
separation---the process through which homogeneous mixtures demix into distinct phases ---has attracted
considerable attention in a variety of fields, incl. physics \cite{Weber_2019,
  rowlinson2013molecular, bray2002theory, doi:10.1063/1.439809,
  doi:10.1063/1.442226, doi:10.1063/1.445747, PhysRevLett.126.028003,
  doi:10.1098/rsif.2021.0255,   PhysRevResearch.3.043150,
  PhysRevLett.125.218003}, mathematics
\cite{aubert2006mathematical, copetti1992numerical,
  gomez2008isogeometric}, chemistry \cite{adamson1967physical,
  ZHAO2015325, zhao2019review, doi:10.1063/1.432687}, material
science \cite{erlebacher2001evolution, chen2002phase,
  Steinbach_2009}, and recently biology \cite{murray2002mathematicalbiology,
  sanders2020competing, hur2020cdk, spoelstra2021crispr}.  
Our understanding of phase separation 
in systems in \cite{PhysRevB.8.1308, cahn1965phase} and out
\cite{bergmann2018active, PhysRevLett.111.145702} of equilibrium is mostly based on mean field
(MF) ideas \cite{safran2018statistical}, also known as regular
solution \cite{cahn1958freeI}, Bragg-Williams \cite{bragg1934effect},
or Flory-Huggins \cite{flory1953principles, doi:10.1063/1.1723621}
theory (for recent works see
\cite{PhysRevLett.126.028003,doi:10.1098/rsif.2021.0255, sanders2020competing,
  PhysRevLett.126.258101, PhysRevResearch.3.043150,
  PhysRevLett.125.268001, PhysRevLett.125.218003, PhysRevX.9.041020,
  hur2020cdk, spoelstra2021crispr, zhang2021decoding,
  PhysRevLett.128.110603, parry2019goldstone}). MF theory 
neglects correlations whose importance grows with the strength of
interactions \cite{Weber_2019}.\ For example, capillary wave fluctuations
  \cite{parry2019goldstone,munster2021interface} are \emph{not}
  captured in MF theories.~This questions whether
MF ideas correctly describe the physics of strongly interacting systems \cite{PhysRevLett.128.038102, zhang2021decoding}. 

Various refined techniques have been developed beyond the MF
approximation, incl. the cavity method \cite{lauber2022statistical},
random phase approximation \cite{lin2018theories, lin2016sequence},
self-consistent field theory \cite{muller2000interface}, and field-theoretic approaches close to criticality
\cite{squarcini2021correlations}. Yet, these techniques either do not
apply to non-uniform systems, or are applicable in a limited range of
interaction strengths. As a result, the phenomenology of phase
separation in the strong-coupling limit remains largely unexplored,
and thus poorly understood.  
\begin{figure}[t!]
    \hspace{-3mm}
    \includegraphics[width = 0.485\textwidth]{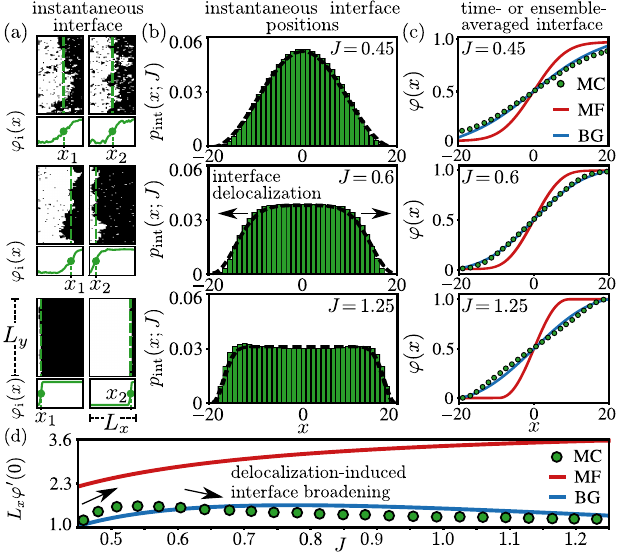}
    \caption{(a)~Realizations of spin configurations
        (top) and corresponding \emph{instantaneous}
        interfaces (bottom) in a two-dimensional Ising strip with lattice constant
        $\delta$ and 
        dimensions $(L_x,L_y){=}(40,120)\delta$, for
        different $J{=}\{0.45,0.6,1.25\}$ obtained
        from Monte-Carlo simulations (see \cite{Note2}
        for
        details).\ The circle and dashed line denote the 
        position 
        of the instantaneous
        interface.\ (b) Statistics 
        of
        interface positions derived from simulations (green) and 
        given by Eq.~\eqref{pdf-interface} (black dashed line) via a mapping onto the Brownian excursion
        problem.\
        (c) Corresponding \emph{ensemble averaged} concentration
        profile along the $x$-axis alongside theoretical predictions
        of mean field (MF; red) and Bethe-Guggenheim (BG; blue) theory.\
        (d) Scaled interface steepness
        $L_x\varphi^{\prime}(0){\equiv} L_x\partial_{x}\varphi(x)|_{x=0}$ as a function of $J$.}
    \label{fig1}
\end{figure}

Here we employ
the Bethe-Guggenheim (BG) approximation
\cite{guggenheim1935statistical, bethe1935statistical,
kikuchi1962theory,doi:10.1063/1.1667895} that includes nearest-neighbor pair-correlations.\ By
evaluating the thermodynamic limit of a spatially inhomogeneous
nearest-neighbor
interacting Ising model, we derive
a Cahn-Hilliard free energy
on the BG level that effectively accounts for the
  effects of capillary
  wave fluctuations.\ We investigate the phenomenology of
interfaces and phase condensation, and find at
sufficiently strong interactions an
  \emph{effective interface broadening} not
  accounted for by MF theories.\ We corroborate the broadening with
  simulations, and
  exact results in the infinite-interaction limit.\ Furthermore, via
numerical simulations of the Cahn-Hilliard equation
\cite{Zwicker2020}, we analyze nucleation kinetics,
and observe amplified nucleation barriers and a non-monotonic dependence of the interface steepness and
critical nucleus size on the interaction strength.
\section{Motivating example:\ Interface delocalization}\label{SecII}
\begin{figure}[t!]
    \includegraphics[width = 0.40\textwidth]{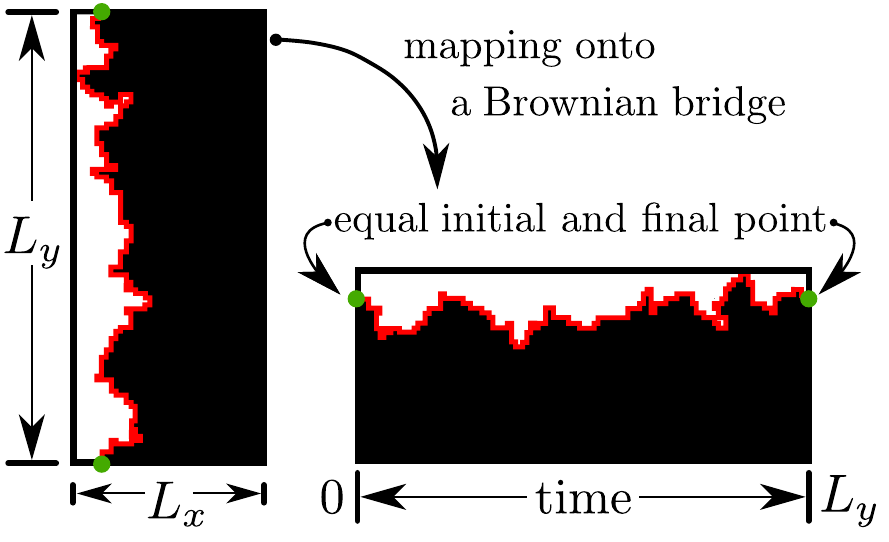}
    \caption{\blue{Mapping the instantaneous interface (red line; left) onto a Brownian bridge (red line; right). Neglecting overhangs, and treating the bulk phases (black/white regions) as homogeneous, the instantenous interface becomes a Brownian trajectory where the vertical coordinate $y$ plays the role of time. Greens dots indicate the equal positions of the interface at $0$ and $L_{y}$, rendering the interface a Brownian bridge.}}
    \label{Fig2}
\end{figure}

An intriguing phenomenon in strongly interacting systems  is interface delocalization \cite{PhysRevE.53.5023, albano1990critical,
    rogiers1993effect, bilalbegovic1988interface, PhysRevLett.74.298,
    binder1999interfacial,
    PhysRevB.34.1932,PhysRevB.47.7519,PhysRevB.49.1092,
    ciach1987scaling,albano2000monte, fisher1984walks}.\ Consider a two-dimensional Ising model with ferromagnetic interaction  $J$ (in units of $k_{\rm B}T$) in a strip geometry (i.e.\ height $\gg$ length)
    in the two-phase regime.\ Imposing periodic boundary conditions in the vertical direction, and thermodynamically co-existing phase compositions at the left/right edges, the \emph{instantaneous} concentration of down-spins  projected onto the $x$ direction, $\varphi_{\rm i}(x)$, develops an  interface (see Fig.~\ref{fig1}a), whose position  $x_{\rm i}$ is defined implicitly via $\varphi_{\rm i}(x_{\rm i})=1/2$.\ In the absence of boundary effects, shifting an instantaneous interface $\varphi_{\rm i}(x_{\rm i})\to\varphi_{\rm i}(x_{\rm i}+{\rm d}x_{\rm i})$ costs no energy.\ However, $x_{\rm i}$ near the  boundaries are entropically penalized, as they allow only for a limited bandwidth of capillary wave fluctuations (see Fig.~\ref{fig1}a, top) \cite{fisher1984walks,albano2000monte,fisher1986interface,PhysRevE.90.012128}.\ As a result, we find at weak to moderate $J$ that the probability density of instantaneous interface positions, defined as $p_{\rm int}(x;J)$, is peaked at the center (see Fig.~\ref{fig1}b, top). At larger $J$ the amplitude of capillary waves diminishes (see Fig.~\ref{fig1}a, center and bottom), and a transition occurs that delocalizes the instantaneous interface (see Fig.~\ref{fig1}b, center and bottom as well as \cite{albano2000monte,PhysRevB.34.1932, fisher1984walks,PhysRevB.49.1092, ciach1987scaling,PhysRevB.47.7519}).\ A sharp but delocalized instantaneous interface becomes effectively broader upon \emph{time- or ensemble-averaging} over respective interface positions (see Fig.~\ref{fig1}c-d). Exact results in the regime $J\to\infty$ have confirmed the interface broadening \cite{PhysRevB.49.1092,PhysRevB.34.1932,ciach1987scaling,PhysRevB.47.7519}, whereas it is known that MF theory fails to account for it \cite{albano2000monte,PhysRevB.27.4499}. A comprehensive theory that captures the broadening transition due to the instantaneous interface delocalization remains elusive.\ This example therefore motivates a deeper and more systematic analysis of interfaces and phase separation in the strong interaction limit.
\blue{
\section{Outline}\label{SecIII}
First, we present in Sec.~\ref{SecIV} a derivation of the probability
density of instantaneous interface positions based on a mapping onto
the Brownian bridge problem (Eq.~\eqref{pdf-interface}).
Thereafter, we present in Sec.~\ref{SecV} a detailed
\blue{microscopic} derivation of the
Cahn-Hilliard\blue{-type} phase-field free energy starting from an
anisotropic two-dimensional Ising model using the BG approximation (Eqs.~\eqref{F}-\eqref{kappa}).
In Sec.~\ref{SecVI} we analyze the field theories by determining the one-dimensional 
equilibrium concentration profile, interface steepness, interface
stiffness, and the critical wavelength of stable perturbations.
Furthermore, we analyze nucleation kinetics via numerical simulations
of the \blue{newly derived} Cahn-Hilliard equation. Finally, in Sec.~\ref{SecVII} we conclude and reflect on possible future directions.}
\blue{
\section{Statistics of instantaneous interface position}\label{SecIV}
Here we derive the probability density of instantaneous interface positions, based on the analogy with Brownian bridges, used for Fig.~\ref{fig1}b (black dashed lines). Furthermore, in Sec.~\ref{SecIVB} we prove the convergence to a uniform
distribution in the limit $J\rightarrow\infty$, which we use in Sec.~\ref{secdisentangle} to disentangle interface delocalization from the instantaneous interface width.
\subsection{Main idea}
Neglecting overhangs, one can map the statistics of instantaneous interfaces onto a one-dimensional confined Brownian bridge problem (see Fig.~\ref{Fig2}) \cite{fisher1984walks}. The idea is to treat the respective bulk phases as ``pure'' (i.e.~homogeneous) and the interface (i.e.~domain wall) as a random walk located between two hard walls at $x=0$ and $x=L_{x}$. Then, in the continuum limit the interface is equivalent to a Brownian trajectory where the vertical coordinate $y$ plays the role of time and the diffusion coefficient is proportional to $1/4\Lambda$ (see Sec.~\ref{SecIVA}), where 
\begin{equation}
    \Lambda=\sinh{(2J+\ln{\tanh{J}})}
    \label{lambda}
\end{equation}
is the exact interface stiffness for the two-dimensional Ising model \cite{PhysRevLett.47.545,PhysRevLett.48.368}. Periodic boundary conditions in the $y$-direction render the trajectories Brownian bridges.}
\blue{
\subsection{Derivation of interface statistics}\label{SecIVA}
We parameterize the domain wall as a Brownian motion $\{x^{\rm i}_y\}_{0 \le y\le L_y}$ where $y$ plays the role of time (or contour length in the polymer context). Then, the Green’s function of the interface with diffusion coefficient $D$ follows the Edwards equation with absorbing boundary conditions at the walls
\begin{align}
    \partial_{y}G(x,y|x_{0})&=D\nabla^{2}_{x}G(x,y|x_{0}), \nonumber \\ G(x,0|x_{0})&=\delta(x-x_{0}), \nonumber \\ G(0,y|x_{0})&=G(L_{x},y|x_{0})=0 , \ \forall \ y\in[0,L_{y}].
    \label{b.c.G.}
\end{align}
The general solution to Eq.~\eqref{b.c.G.} is
\begin{equation*}
    G(x,y|x_{0})=\frac{2}{L_{x}}\sum_{k=1}^{\infty}\sin{\left(\frac{k\pi x}{L_{x}}\right)}\sin{\left(\frac{k\pi x_{0}}{L_{x}}\right)}\e{\frac{-\pi^{2}k^{2}Dy}{L^{2}_{x}}}.
\end{equation*}
Particularly interesting is the mean squared displacement, which 
for $y\ll D/L^{2}_{x}$ is given by $\langle (x{-}x_{0})^2 \rangle {\simeq} 2Dy$, where $\simeq$ stands for asymptotic equality, i.e. $A\simeq B$ when $A/B\to 1$. Such a scaling is expected for a freely diffusing particle. Now we recall the exact results of Abraham \cite{PhysRevLett.47.545} and Fisher \cite{PhysRevLett.48.368} who found that for the two-dimensional Ising strip the interface width should scale as $\langle (x-x_{0})^2 \rangle {\propto} y/\Lambda$ where the proportionality factor includes some lattice length scale and $\Lambda$ is the surface stiffness given by Eq.~\eqref{lambda}. This outcome allows us to relate the diffusion coefficient $D$ to the surface stiffness 
\begin{equation}
    D\propto\frac{1}{2\Lambda}.
    \label{D-J-relation}
\end{equation}
Under periodic boundary conditions in the $y$-direction the interfaces have an equal position at $y=0$ and $y=L_{y}$, also known as Brownian bridges. In this case the propagator is simply given by $G(x,L_{y}|x)$. We can now calculate the probability density to have an interface located at position $x$, which upon normalization is given by
\begin{align}
    p_{\rm int}(x;J)&=\frac{G(x,L_{y}|x)}{\int_{0}^{L_{x}} G(x,L_{y}|x)dx}
    \nonumber\\
    &=\frac{1}{L_{x}}\frac{\vartheta_{3}(0,\e{-\alpha_{J}})-\vartheta_{3}(\pi x /L_{x},\e{-\alpha_{J}})}{\vartheta_{3}(0,\e{-\alpha_{J}})-1},
    \label{pdf-interface}
\end{align}
where $\alpha_{J}\equiv \pi^{2}DL_{y}/L^{2}_{x}$ and
$\vartheta_{3}(a,x)$ is Jacobi's elliptic theta of the third kind. The second equality can be obtained from the definition of $\vartheta_{3}(a,x)$. Plugging Eq.~\eqref{D-J-relation} for the diffusion coefficient (with proportionality factor equal to unity) into Eq.~\eqref{pdf-interface}, we obtain the black dashed lines in Fig.~\ref{fig1}b. As long as $J\ll1$ we have $\Lambda \ll 1$ and thus $\alpha_{J}\gg1$.\ Accordingly, Eq.~\eqref{pdf-interface} predicts instantaneous
interfaces to be localized with a probability density $p_{\rm
int}(x;J)\propto \sin(\pi x/L_x)^2$ (see Fig.~\ref{fig1}b, top panel).\ Conversely, for sufficiently large $J$ we find $\Lambda \gg 1$ and hence $\alpha_{J}\ll 1$, and the interface positions become delocalized (see Fig.~\ref{fig1}b, center and bottom panels).}
\blue{
\subsection{Convergence to the uniform distribution}\label{SecIVB}
Next, we show that Eq.~\eqref{pdf-interface} converges to
a uniform distribution for $J{\rightarrow}\infty$. First, we define $q{\equiv}\mathrm{e}^{-\alpha_{J}}$ and rewrite
\begin{align}
    \vartheta_{3}(\pi
    x/L_{x},\mathrm{e}^{-\alpha_{J}})
    =\sum_{n=-\infty}^{\infty}q^{n^2}\left( \mathrm{e}^{\mathrm{i}2\pi x/L_x}\right)^n.
    \label{Jac}
\end{align}
Since $\alpha_{J}\ge 0$ and $\lim_{J\to\infty}\alpha_{J}=0$ (uniformly), we find that $\lim_{J\to\infty}\mathrm{e}^{-\alpha_{J}}$ is equivalent to $\lim_{q\uparrow 1}q$ in Eq.~\eqref{Jac}. We now use the asymptotic result for $q\uparrow
1$ \cite{Wang}
\begin{equation}
    \lim\limits_{q\uparrow1}\sum_{n=-\infty}^{\infty}q^{n^2}\left( \mathrm{e}^{\mathrm{i}2\pi
    x/L_x}\right)^n\simeq \sqrt{\frac{\pi}{-\ln
    q}}\exp\left(\frac{\pi^2x^2}{L_x^2\ln q}\right).
    \label{asym}  
\end{equation}
Let us now rewrite Eq.~\eqref{pdf-interface} as
\begin{align}
    L_{x}p_{\rm int}(x;J)&=\frac{\vartheta_{3}(0,\e{-\alpha_{J}}){-}1{-}\vartheta_{3}(\pi
    x /L_{x},\e{-\alpha_{J}}){+}1}{\vartheta_{3}(0,\e{-\alpha_{J}})-1}\nonumber\\
    &=1-\frac{\vartheta_{3}(\pi x /L_{x},\e{-\alpha_{J}})-1}{\vartheta_{3}(0,\e{-\alpha_{J}})-1}.
    \label{rew}
\end{align}
We can now evaluate the limit of Eq.~\eqref{rew} using Eq.~\eqref{asym}. Note that Eq.~\eqref{asym}$\ \gg 1$ for $0<x<L_x$. Hence, we find 
\begin{align}
    \lim_ {J\to\infty}L_{x}p_{\rm int}(x;J)&\simeq
    1-\lim_{q\uparrow 1}\exp\left(\frac{\pi^2x^2}{L_x^2\ln
    q}\right)\nonumber\\
    &=1-\lim_{\alpha_{J}\downarrow 0}\exp\left(-\frac{\pi^2x^2}{L_x^2\alpha_{J}}\right)\to 1,
\end{align}
for $0{<}x{<}L_x$, while we have $p_{\rm int}(0;J){=}p_{\rm int}(L_x;J){=}0 , \ \forall J$. In the forthcoming sections we take the boundaries at $x=\pm L_{x}/2$, which  shifts the coordinates to $x{\rightarrow} x-L_{x}/2$.}

\blue{Notably, when $J\to\infty$ a Casimir effect appears in addition (see e.g.\ \cite{Casimir,Casimir2}) that is \emph{not} captured in Eq.~\eqref{pdf-interface}, i.e.\ the entropy due to \emph{bulk} fluctuations is enhanced near the boundaries resulting in ``peaks'' (see Fig.~\ref{fig1}b, bottom).}
\begin{figure}[t!]
    \includegraphics[width = 0.48\textwidth]{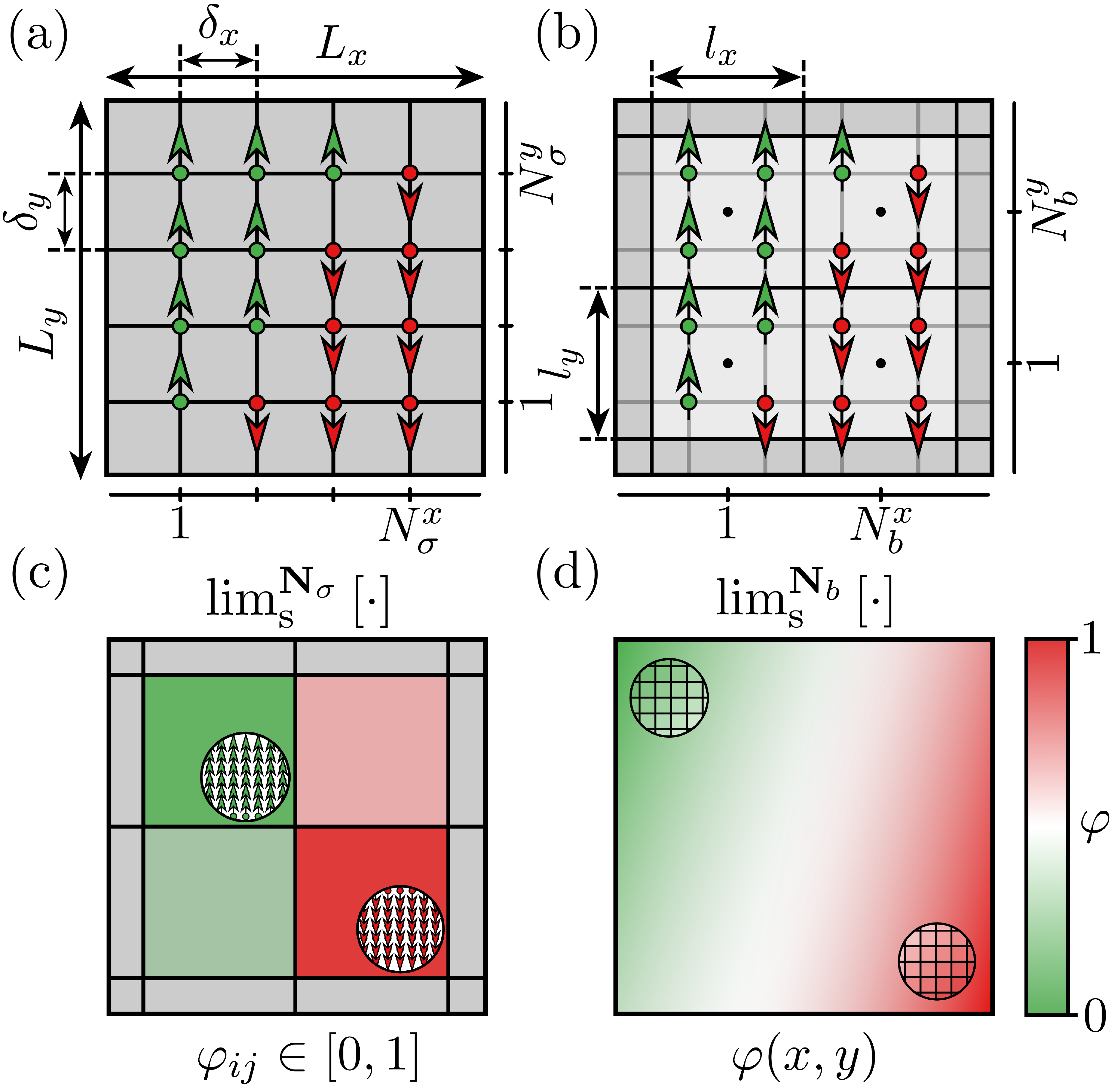}
    \caption{\blue{(a-b) Lattice setup of the spins (a) and spin blocks (b).~$(L_{x,y},l_{x,y},\delta_{x,y})$ are the lattice length, spin block length, and lattice spacing, respectively.~The number of spins and spin blocks are denoted with $(N^{x,y}_{\sigma},N^{x,y}_{b})$.~Here we consider an example with sixteen spins and four spin blocks. (c-d) Thermodynamic limit of the spins (c) and spin blocks (d) defined in Eq.~\eqref{TDlims}.~The circles display a graphical magnification of individual spins (c) and spin blocks (d).}} 
    \label{Fig3}
\end{figure}
\blue{
\section{Cahn-Hilliard free energy including pair correlations}\label{SecV}
\subsection{Lattice setup and the thermodynamic limit}
\textit{Spins.---}For simplicity, and without much loss
of generality, we limit the discussion
to two-dimensional systems with horizontal and vertical direction $\mathbf{x}{=}(x,y){\in}\mathbb{R}^{2}$, respectively. We consider $N_{\sigma}=N_{\sigma}^{x}\times N_{\sigma}^{y}$ spins $\sigma_{ij}=\pm 1$ with $(i,j)\in(\{1,...,N^{x}_{\sigma}\},\{1,...,N^{y}_{\sigma}\})$ arranged
on a lattice with sides $(L_{x},L_{y})$. In Fig.~\ref{Fig3}a we provide an example of a square lattice with sixteen spins. The lattice spacings between spins are $(\delta_{x},\delta_{y}){=}(L_x/N^{x}_{\sigma},L_y/N^{y}_{\sigma})$. The lattice coordination number is denoted with $z{=}z_{x}{+}z_{y}$, and
$\mathbf{z}{=}{\rm diag}(z_{x},z_{y})$ is a diagonal matrix containing the lattice coordination numbers in each direction. The square lattice in Fig.~\ref{Fig3} has $(z_{x},z_{y})=(2,2)$ and $z=4$.}

\blue{\textit{Spin blocks.---}Similar to Kadanoff's block spin method \cite{PhysicsPhysiqueFizika.2.263} we place spins into $N_{b}=N^{x}_{b}\times N^{y}_{b}$ blocks as shown in Fig.~\ref{Fig3}b. Let $\mathbf{b}_{ij}$ with $(i,j)\in(\{1,...,N^{x}_{b}\},\{1,...,N^{y}_{b}\})$ denote such a block containing $N^{b}_{\sigma}=N_{\sigma}/N_{b}$ spins. 
Consequently, the horizontal and vertical length of each block is given by $(l_{x},l_{y})=(L_{x}/N^{x}_{b},L_{y}/N^{y}_{b})$. The blocks have the same lattice coordination number as the spins. In Fig.~\ref{Fig3}b each block has four spins and aligns with 2 blocks in the horizontal and vertical direction, respectively.}

\blue{\textit{Thermodynamic limit.---}To construct a Cahn-Hilliard free energy we introduce the following two scaling limits where we take the number of spins/blocks to infinity while simultaneously keeping the block/lattice length fixed, i.e.\
\begin{eqnarray}    
    {\rm lim}^{\mathbf{N}_{\sigma}}_{\rm s}\left[\cdot\right]&\equiv&\textstyle{\lim^{N^{x}_{\sigma},N^{y}_{\sigma}\rightarrow\infty}_{l_{x},l_{y}={\rm const.}}}\left[\cdot\right], \nonumber \\ 
    {\rm lim}^{\mathbf{N}_{b}}_{\rm s}\left[\cdot\right]&\equiv&\textstyle{\lim^{N^{x}_{b},N^{y}_{b}\rightarrow\infty}_{L_{x},L_{y}={\rm const.}}}\left[\cdot\right],
    \label{TDlims}
\end{eqnarray} 
where $\mathbf{N}_{\sigma}$ and $\mathbf{N}_{b}$ denote the
thermodynamic limit of the spins and blocks,
respectively. In Fig.~\ref{Fig3}c-d we give a schematic representation of both limits.
\subsection{Coarse-grained lattice observables}
\textit{Fraction of down spins.---}The fraction of down spins in block $\mathbf{b}_{ij}$ (containing $N^{b}_{\sigma}$ spins) is defined as 
\begin{equation} 
    \varphi_{ij}(\{\mathbf{b}_{ij}\})\equiv(N^{b}_{\sigma})^{-1}\smashoperator{\sum_{mn\in \mathbf{b}_{ij}}}(1-\sigma_{mn})/2,
    \label{phi}
\end{equation}
where $mn\in\mathbf{b}_{ij}$ denotes a sum over all indices within block $\mathbf{b}_{ij}$. For a finite number of spins within each block $\varphi_{ij}$ is a rational number. Applying the first scaling limit in Eq.~\eqref{TDlims} takes the number of spins within each block to infinity, rendering ${\rm lim}^{\mathbf{N}_{\sigma}}_{\rm s}\left[\varphi_{ij}\right]\in[0,1]$ a continuous variable (see Fig.~\ref{Fig3}c). The second limit takes the number of blocks to infinity while simultaneously decreasing their distance, resulting in a continuous differentiable field 
\begin{equation}
    {\rm lim}^{\mathbf{N}_{b}}_{\rm s}[{\rm lim}^{\mathbf{N}_{\sigma}}_{\rm s}\left[\varphi_{ij}\right]]=\varphi(\mathbf{x}) \nonumber,
\end{equation}
as depicted in Fig.~\ref{Fig3}d.}

\blue{\textit{Intrablock defects.---}Additionally to Eq.~\eqref{phi}, we need to define the fraction of intrablock defects inside $\mathbf{b}_{ij}$ in the horizontal and vertical direction, which are given by
\begin{align}
    \zeta^{x,y}_{ij}(\{\mathbf{b}_{ij}\})&\equiv(z_{x,y}N^{b}_{\sigma})^{-1}\smashoperator{\sum_{\langle mn,kl \rangle_{x,y}\in \mathbf{b}_{ij}}}|\sigma_{mn}{-}\sigma_{kl}|,
    \label{zeta-y}
\end{align}
where $\langle mn,kl \rangle_{x,y}\in\mathbf{b}_{ij}$ denotes nearest neighbors within block $\mathbf{b}_{ij}$ in the horizontal ($x$) and vertical ($y$) direction, respectively.} 

\blue{\textit{Interblock defects.---}Finally, we define the fraction of interblock defects between neighbouring blocks in the horizontal and vertical direction, respectively given by
\begin{align}
    \xi^{x\pm}_{ij}(\{\mathbf{b}_{ij},\mathbf{b}_{i\pm1j}\})&\equiv(z_{x}N^{b}_{\sigma})^{-1}\smashoperator{\sum_{\langle mn,kl \rangle_{x}\in (\mathbf{b}_{ij},\mathbf{b}_{i\pm1j})}}|\sigma_{mn}-\sigma_{kl}|,\nonumber\\
    \xi^{y\pm}_{ij}(\{\mathbf{b}_{ij},\mathbf{b}_{ij\pm1}\})&\equiv(z_{y}N^{b}_{\sigma})^{-1}\smashoperator{\sum_{\langle mn,kl\rangle_{y}\in (\mathbf{b}_{ij},\mathbf{b}_{ij\pm1})}}|\sigma_{mn}-\sigma_{kl}|,
    \label{xi-y}
\end{align}
where one needs to account for the boundary conditions upon summing over boundary blocks.}

\blue{As an example, the fraction of down spins, intra- and interblock defects for the lower left block in Fig.~\ref{Fig3}b under periodic boundary conditions is: $\varphi_{11}=1/4$, $\zeta^{x,y}_{11}=1/8$, $\xi^{x+}_{11}=1/8$, $\xi_{11}^{x-}=1/4$, $\xi_{11}^{y+}=0$, and $\xi_{11}^{y-}=1/8$.
\subsection{Coarse-grained Ising Hamiltonian}
We now rewrite the nearest neighbour Ising Hamiltonian in terms of the coarse-grained intensive lattice observables introduced in Eqs.~\eqref{zeta-y}-\eqref{xi-y}. The Ising Hamiltonian reads (in units of $k_{\rm B}T$)
\begin{equation}
    \mathcal{H}(\boldsymbol{\sigma})=-J_{x}\smashoperator{\sum\limits_{\langle mn,kl \rangle_{x}}}\sigma_{mn}\sigma_{kl}-J_{y}\smashoperator{\sum\limits_{\langle mn,kl \rangle_{y}}}\sigma_{mn}\sigma_{kl},
    \label{H}
\end{equation}
where $\boldsymbol{\sigma}$ is the matrix containing all spin
configurations, and $J_{x,y}\geq0$ is the
ferromagnetic interaction strength in the horizontal and vertical
direction, respectively. To make use of
Eqs.~\eqref{zeta-y}-\eqref{xi-y} we insert the 
identities
\begin{align*}
    \sigma_{mn}\sigma_{kl}&=1-|\sigma_{mn}-\sigma_{kl}|, \\
    \smashoperator{\sum\limits_{\langle mn,kl \rangle_{x,y}}} 1 &= z_{x,y}N^{x,y}_{\sigma}/2,
\end{align*}
(the latter can also be read as a definition for $z_{x,y}$) and decompose the Hamiltonian into a sum over inter- and intrablock contributions 
\begin{equation}
    \mathcal{H}(\boldsymbol{\sigma})=N^{b}_{\sigma}\sum_{i=1}^{N^{x}_{b}}\sum_{j=1}^{N^{y}_{b}}
    [\mathcal{H}_{\rm inter}+\mathcal{H}_{\rm intra}-\mathcal{C}].
    \label{H2}
\end{equation}
The respective contributions inside the sum are given by
\begin{align}
    \mathcal{H}_{\rm inter}(\{\mathbf{b}_{i\pm1j},\mathbf{b}_{ij},\mathbf{b}_{ij\pm1}\})&=z_{x} J_{x}(\xi^{x+}_{ij}{+}\xi^{x-}_{ij})/2{+}(x{\leftrightarrow}y),
    \nonumber\\
    \mathcal{H}_{\rm intra}(\{\mathbf{b}_{ij}\})&=z_{x}J_{x}\zeta^{x}_{ij}+(x\leftrightarrow y),\nonumber\\
     \mathcal{C}&=(z_{x}J_{x}+z_{y}J_{y})/2.
    \label{Hinter}
\end{align}
The term
$(x\leftrightarrow y)$ in Eq.~\eqref{Hinter} denotes
a repetition of the preceding term with $x$ and $y$
interchanged. Eq.~\eqref{H2} is an \emph{exact} expression for the Ising Hamiltonian in terms of spin blocks. Note that the factor $1/2$ in $\mathcal{H}_{\rm inter}$ accounts for the double counting over interblock contributions. 
\subsection{Coarse-grained partition function}
Since the Hamiltonian is decomposed into a sum over spin blocks, the partition function $Z$ can be factorized into a product of partition functions per block $\mathcal{Z}_{ij}$. Inserting Eq.~\eqref{H2} into the partition function leads to the following \emph{exact} expression
\begin{align}
    Z{=}\prod_{i=1}^{N^{x}_{b}}\prod_{j=1}^{N^{y}_{b}}\mathcal{Z}_{ij}{\equiv}\prod_{i=1}^{N^{x}_{b}}\prod_{j=1}^{N^{y}_{b}}\sum_{\{\mathbf{b}_{ij}\}}\e{-N^{b}_{\sigma}[\mathcal{H}_{\rm inter}+\mathcal{H}_{\rm intra}-\mathcal{C}]}.
    \label{Z1}
\end{align}
The evaluation of the configurational sum over all possible spin block configurations $\{\mathbf{b}_{ij}\}$ constitutes a difficult --- if not impossible --- task. However, the Hamiltonian given by Eq.~\eqref{Hinter} solely depends on the seven lattice observables $(\varphi_{ij},\zeta^{x,y}_{ij},\xi^{x,y\pm}_{ij})$. Therefore, we can interchange the configurational sum by a sum over all possible values of these seven lattice observables, and introduce a \emph{degeneracy of states} --- $\Psi(\varphi_{ij},\zeta^{x,y}_{ij},\xi^{x,y\pm}_{ij})$ --- which accounts for the multiplicity of configurations. Then we obtain
\begin{equation}
    \mathcal{Z}_{ij}{=}\sum_{\varphi_{ij}}\smashoperator{\sum_{\zeta^{x,y}_{ij}}}\smashoperator{\sum_{\xi^{x,y\pm}_{ij}}}\Psi(\varphi_{ij},\zeta^{x,y}_{ij},\xi^{x,y\pm}_{ij})
    \e{{-}N^{b}_{\sigma}[\mathcal{H}_{\rm inter}+\mathcal{H}_{\rm intra}-\mathcal{C}]}.
    \label{Z}
\end{equation}
Equation~\eqref{Z} is an exact expression as long as the degeneracy of states  $\Psi(\varphi_{ij},\zeta^{x,y}_{ij},\xi^{x,y\pm}_{ij})$ is evaluated exactly.}

\blue{\textit{Normalization condition.---}For $(J_{x},J_{y})=(0,0)$ the degeneracy of states should obey the following relation
\begin{equation}
    \sum_{\zeta^{x,y}_{ij}}\sum_{\xi^{x,y\pm}_{ij}}\Psi(\varphi_{ij},\zeta^{x,y}_{ij},\xi^{x,y\pm}_{ij})\stackrel{!}{=}\binom{N^{b}_{\sigma}}{\varphi_{ij} N^{b}_{\sigma}},
    \label{norm}
\end{equation}
since this is the number of possible configurations to place
$\varphi_{ij} N^{b}_{\sigma}$ down spins in a block that contains
$N^{b}_{\sigma}$ spins in total. We will use Eq.~\eqref{norm} as a normalization condition to consistently approximate the
degeneracy of states. 
\subsection{Pair-approximation Ansatz}
\textit{Intuition behind the BG approximation.---}Our next aim is to approximate the degeneracy of states \emph{by placing spin pairs onto the lattice}. Imagine that we are given a number of spin pairs with $N_{\uparrow\uparrow}$, $N_{\downarrow\downarrow}$, and $N_{\uparrow\downarrow}$ denoting the number of up-up, down-down, and up-down (i.e.\ defects) spin pairs. The total number of distinct lattice configurations for fixed $(N_{\uparrow\uparrow}, N_{\downarrow\downarrow}, N_{\uparrow\downarrow})$ is given by \cite{PhysRev.81.988}
\begin{equation*}
    \Psi\approx\Psi_{\rm BG}\equiv\frac{(N_{\uparrow\uparrow}+N_{\uparrow\downarrow}+N_{\downarrow\downarrow})!}{(N_{\uparrow\uparrow})!(N_{\uparrow\downarrow}/2)!^{2}(N_{\downarrow\downarrow})!},
\end{equation*}
where the factor $1/2$ in the denominator accounts for the symmetry $N_{\uparrow\downarrow}=N_{\downarrow\uparrow}$.
For even $N_{\uparrow\downarrow}$ the term $(N_{\uparrow\downarrow}/2)!$ is well-defined. However, when $N_{\uparrow\downarrow}$ is odd we are forced to consider the generalized factorial
\begin{equation}
    \Psi\approx\Psi_{\rm BG}\equiv\frac{\Gamma(N_{\uparrow\uparrow}+N_{\uparrow\downarrow}+N_{\downarrow\downarrow}+1)}{\Gamma(N_{\uparrow\uparrow}+1)\Gamma(N_{\uparrow\downarrow}/2+1)^{2}\Gamma(N_{\downarrow\downarrow}+1)},
    \label{PsiBG2}
\end{equation}
where $\Gamma(x)$ is the Gamma function \cite{artin2015gamma}.  
Equation \eqref{PsiBG2} comprises the main essence of the BG approximation \footnote{Equation \eqref{PsiBG2} is exact when there are no closed loops in the lattice. Therefore the BG approximation is exact on the Bethe lattice.}.}

\blue{\textit{Non-uniform degeneracy of states.---}To account for
a non-uniform concentration profile we need to construct the
degeneracy of states for each of the
invidual blocks $\mathbf{b}_{ij}$. Similar to Eq.~\eqref{PsiBG2} we
introduce a pair-approximation Ansatz for the degeneracy of
states. The difference, however, is that we must now distinguish between intra- and interblock contributions. Furthermore, we want to express the degeneracy of states in terms of Eqs.~\eqref{phi}-\eqref{xi-y}. This results in 
\begin{equation}
    \Psi(\varphi_{ij},\zeta^{x,y}_{ij},\xi^{x,y\pm}_{ij})\approx
    \mathcal{N}(\varphi_{ij})\hat{\Psi}(\varphi_{ij},\zeta^{x,y}_{ij},\xi^{x,y\pm}_{ij}),
    \label{Psi}
\end{equation}
where $\mathcal{N}(\varphi_{ij})$ is a normalization constant left to be determined. The unnormalized degeneracy of states reads
\begin{equation}
    \hat{\Psi}\equiv
    \hat{\Psi}^{-\frac{1}{2}}_{\rm intra}(\varphi_{ij},\zeta^{x,y}_{ij})\prod_{\pm}\hat{\Psi}^{-\frac{1}{4}}_{\rm inter}(\varphi_{ij},\xi^{x,y\pm}_{ij}),
    \label{Psi-hat}
\end{equation}
which is divided into intra- and interblock contributions
\begin{align}
    \hat{\Psi}_{\rm intra}&\equiv
    \psi_{x}(\varphi_{ij},\varphi_{ij},\zeta^{x}_{ij})\psi_{y}(\varphi_{ij},\varphi_{ij},\zeta^{y}_{ij})
    \label{Psi-hat-intra},\\
    \hat{\Psi}_{\rm inter}&\equiv\psi_{x}(\varphi_{ij},\varphi_{i\pm1j},\xi^{x\pm}_{ij})\psi_{y}(\varphi_{ij},\varphi_{ij\pm1},\xi^{y\pm}_{ij}),
    \label{Psi-hat-inter}
\end{align}
and the auxiliary functions $\psi_{x,y}(a,b,c)$ are given by
\begin{equation*}
    \psi_{x,y}(a,b,c){\equiv}
    \hat{\Gamma}_{x,y}(1-a-c)\hat{\Gamma}_{x,y}(b-c)\hat{\Gamma}_{x,y}(a-b+c)\hat{\Gamma}_{x,y}(c),
\end{equation*}
with $\hat{\Gamma}_{x,y}(w)\equiv \Gamma(z_{x,y}N^{b}_{\sigma}w/2+1)$,
and $\Gamma(w)$ being the Gamma function. Equation~\eqref{Psi} can be derived similarly to Eq.~\eqref{PsiBG2} by counting the number of degenerate configurations upon distributing spin pairs over a lattice. The functions $\psi_{x,y}(\varphi_{ij},\varphi_{ij},\zeta^{x,y}_{ij})$ account for distributing spin pairs inside a single block in the horizontal and vertical direction, respectively. Similarly, $\psi_{x}(\varphi_{ij},\varphi_{i\pm1j},\xi^{x\pm}_{ij})$ and $\psi_{y}(\varphi_{ij},\varphi_{ij\pm1},\xi^{y\pm}_{ij})$ account for distributing spin pairs between two neighbouring blocks in the horizontal and vertical direction, respectively.
\subsection{Evaluation of normalization constant}
The normalization constant $\mathcal{N}(\varphi_{ij})$ in Eq.~\eqref{Psi} is determined by equation Eq.~\eqref{norm}. To evaluate the six sums over the lattice observables $(\zeta^{x,y}_{ij},\xi^{x,y\pm}_{ij})$ we take the thermodynamic limit of the spins, rendering the observables continuous, and employ the maximum term method (i.e.\ saddle point approximation). This gives the following maximizing arguments (henceforth indicated with a hat)
\begin{widetext}
    \begin{align}
    \hat{\zeta}^{x,y}(\varphi_{ij})&\equiv\arg\sup\nolimits_{\zeta^{x,y}_{ij}}\{{\rm lim}^{\mathbf{N}_{\sigma}}_{\rm s}[(N^{b}_{\sigma})^{-1}\ln{(\psi_{x,y}^{-\frac{1}{2}}(\varphi_{ij},\varphi_{ij},\zeta^{x,y}_{ij})})]\}=\varphi_{ij}(1-\varphi_{ij}), \nonumber \\  \hat{\xi}^{x\pm}(\varphi_{i\pm1j},\varphi_{ij})&\equiv\arg\sup\nolimits_{\xi^{x\pm}_{ij}}\{{\rm lim}^{\mathbf{N}_{\sigma}}_{\rm s}[(N^{b}_{\sigma})^{-1}\ln{(\psi_{x}^{-\frac{1}{4}}(\varphi_{ij},\varphi_{i\pm1j},\xi^{x\pm}_{ij}))}]\}=\varphi_{i\pm1j}(1-\varphi_{ij}), \nonumber \\
    \hat{\xi}^{y\pm}(\varphi_{ij\pm1},\varphi_{ij})&\equiv\arg\sup\nolimits_{\xi^{y\pm}_{ij}}\{{\rm lim}^{\mathbf{N}_{\sigma}}_{\rm s}[(N^{b}_{\sigma})^{-1}\ln{(\psi_{y}^{-\frac{1}{4}}(\varphi_{ij},\varphi_{ij\pm1},\xi^{y\pm}_{ij}))}]\}=\varphi_{ij\pm1}(1-\varphi_{ij}).
    \label{zeta-xi-norm}
    \end{align}
\end{widetext}
To obtain Eq.~\eqref{zeta-xi-norm} we used Stirling's
approximation for the Gamma function
$\ln{\Gamma(x)}=\Xi(x)-x+\mathcal{O}(\ln{x})$  for ${\rm Re}(x)>0$
with $\Xi(x)\equiv x\ln{(x)}$.  Plugging Eq.~\eqref{zeta-xi-norm}
into Eq.~\eqref{Psi}, and finally solving Eq.~\eqref{norm} for the
normalization  constant yields 
\begin{equation*}
    \mathcal{N}(\varphi_{ij})=
    \binom{N^{b}_{\sigma}}{\varphi_{ij}N^{b}_{\sigma}}\hat{\Psi}^{-1}(\varphi_{ij},\hat{\zeta}^{x,y},\hat{\xi}^{x,y\pm}).
\end{equation*}
\subsection{Evaluation of partition function}
With the normalization constant evaluated explicitly, we can now determine the partition function given by Eq.~\eqref{Z} in combination with \eqref{Psi}. Again, we take the thermodynamic limit of the spins and approximate the six inner sums over $(\zeta^{x,y}_{ij},\xi^{x,y\pm}_{ij})$ in Eq.~\eqref{Z} with
the maximum term method, giving the following extremizing arguments (henceforth indicated with a hat + dagger)
\begin{widetext}
    \begin{align}
    \hat{\zeta}^{x,y}_{\dagger}(\varphi_{ij})
    &\equiv\arg\sup\nolimits_{\zeta^{x,y}_{ij}}\{{\rm lim}^{\mathbf{N}_{\sigma}}_{\rm s}[(N^{b}_{\sigma})^{-1}\ln{(\psi_{x,y}^{-\frac{1}{2}}(\varphi_{ij},\varphi_{ij},\zeta^{x,y}_{ij})\e{-z_{x,y}J_{x,y}\zeta^{x,y}_{ij}N^{b}_{\sigma}}
    )}]\}
    =2\hat{\zeta}^{x,y}/\Omega_{x,y}(0,\hat{\zeta}^{x,y}),\label{zeta-xi-par3}\\
    \hat{\xi}^{x\pm}_{\dagger}(\varphi_{i\pm1j},\varphi_{ij})&\equiv\arg\sup\nolimits_{\xi^{x\pm}_{ij}}\{{\rm lim}^{\mathbf{N}_{\sigma}}_{\rm s}[(N^{b}_{\sigma})^{-1}\ln{(\psi^{-\frac{1}{4}}_{x}(\varphi_{ij},\varphi_{i\pm1j},\xi^{x\pm}_{ij})\e{-z_{x}J_{x}\xi^{x\pm}_{ij}N^{b}_{\sigma}/2})}]\}
    =2\hat{\xi}^{x\pm}/\Omega_{x}(\varphi_{ij}{-}\varphi_{i\pm1j},\hat{\xi}^{x\pm}),\nonumber\\
    \hat{\xi}^{y\pm}_{\dagger}(\varphi_{ij\pm1},\varphi_{ij})&\equiv\arg\sup\nolimits_{\xi^{y\pm}_{ij}}\{{\rm lim}^{\mathbf{N}_{\sigma}}_{\rm s}[(N^{b}_{\sigma})^{-1}\ln{(\psi^{-\frac{1}{4}}_{y}(\varphi_{ij},\varphi_{ij\pm1},\xi^{y\pm}_{ij})\e{-z_{y}J_{y}\xi^{y\pm}_{ij}N^{b}_{\sigma}/2})}]\}=2\hat{\xi}^{y\pm}/\Omega_{y}(\varphi_{ij}{-}\varphi_{ij\pm1},\hat{\xi}^{y\pm})\nonumber,
    \end{align}
\end{widetext}
where we introduced the auxiliary function
\begin{equation*}
    \Omega_{x,y}(a,b){\equiv}1+a\gamma_{x,y}+[\delta_{a,0}{+}{\rm sgn}(a)]([1+a\gamma_{x,y}]^2+4b\gamma_{x,y})^{\frac{1}{2}},
\end{equation*}
with ${\rm sgn}(x){=}\pm 1$ for $\pm x{>}0$, ${\rm sgn}(0){=}0$, and 
$\gamma_{x,y}{\equiv}{\rm e}^{4J_{x,y}}{-}1$. For
$(J_{x},J_{y}){=}(0,0)$ we have $\hat{\zeta}^{x,y}_{\dagger}{=}\hat{\zeta}^{x,y}$ and $\hat{\xi}^{x,y\pm}_{\dagger}{=}\hat{\xi}^{x,y\pm}$, as
expected from their definition. With the six inner sums in Eq.~\eqref{Z} evaluated, we are left with the sum over $\varphi_{ij}$. To evaluate the last sum we introduce the \emph{free energy density} in the thermodynamic limit of the spins
\begin{widetext}
\begin{align}
    &{\rm f}(\varphi_{i\pm1j},\varphi_{ij},\varphi_{ij\pm1})\equiv{\rm lim}^{\mathbf{N}_{\sigma}}_{\rm s}\left[-(N^{b}_{\sigma})^{-1}\ln{(\Psi(\varphi_{ij},\hat{\zeta}^{x,y}_{\dagger},\hat{\xi}^{x,y\pm}_{\dagger})\e{-N^{b}_{\sigma}[z_{x}J_{x}(\hat{\zeta}^{x}_{\dagger}+\{\hat{\xi}^{x+}_{\dagger}+\hat{\xi}^{x-}_{\dagger}\}/2)+(x\leftrightarrow y)-\mathcal{C}]})}\right] \nonumber \\
    &=(z_{x}/8)\textstyle{\sum_{\pm}}[\Xi(1-\varphi_{ij}-\hat{\xi}^{x\pm}_{\dagger})+\Xi(\varphi_{i\pm1j}-\hat{\xi}^{x\pm}_{\dagger})+\Xi(\varphi_{ij}-\varphi_{i\pm1j}+\hat{\xi}^{x\pm}_{\dagger})+\Xi(\hat{\xi}^{x\pm}_{\dagger})-\Xi(1-\varphi_{i\pm1j})-\Xi(\varphi_{i\pm1j})]\nonumber\\
    &+(z_{y}/8)\textstyle{\sum_{\pm}}[\Xi(1-\varphi_{ij}-\hat{\xi}^{y\pm}_{\dagger})+\Xi(\varphi_{ij\pm1}-\hat{\xi}^{y\pm}_{\dagger})+\Xi(\varphi_{ij}-\varphi_{ij\pm1}+\hat{\xi}^{y\pm}_{\dagger})+\Xi(\hat{\xi}^{y\pm}_{\dagger})-\Xi(1-\varphi_{ij\pm1})-\Xi(\varphi_{ij\pm1})]\nonumber\\
    &+\{(z_{x}/4)[\Xi(1-\varphi_{ij}-\hat{\zeta}^{x}_{\dagger})+\Xi(\varphi_{ij}-\hat{\zeta}^{x}_{\dagger})+2\Xi(\hat{\zeta}^{x}_{\dagger})]+(x\leftrightarrow y)\}+(1-3z/4)[\Xi(\varphi_{ij})+\Xi(1-\varphi_{ij})]\nonumber\\
    &+\{z_{x}J_{x}(\hat{\zeta}^{x}_{\dagger}+\{\hat{\xi}^{x+}_{\dagger}+\hat{\xi}^{x-}_{\dagger}\}/2)+(x\leftrightarrow y)\}-\mathcal{C},
    \label{fj_BG}
\end{align}
\end{widetext}
where $(x\leftrightarrow y)$ always applies directly to its preceding term, and we recall that $\Xi(x)\equiv x\ln{(x)}$. To optimize Eq.~\eqref{fj_BG} over $\varphi_{ij}$, we can employ two different strategies: 
\begin{enumerate}
    \item Optimize ${\rm f}(\varphi_{i\pm1j},\varphi_{ij},\varphi_{ij\pm1})$ and apply ${\rm lim}^{\mathbf{N}_{b}}_{\rm s}\left[\cdot\right]$.
    \item Take ${\rm lim}^{\mathbf{N}_{b}}_{\rm s}[{\rm f}(\varphi_{i\pm1j},\varphi_{ij},\varphi_{ij\pm1})]$ and then optimize.
\end{enumerate}
In \footnote{See Supplementary Material at [...]} we apply the first strategy, and here we proceed with the second. In evaluating the ther modynamic limit of the blocks in Eq.~\eqref{fj_BG} we need to keep track of various terms, which is done explicitly in \cite{Note2}. Here we fast-forward to the final result. Restoring the product over the spin blocks in Eq.~\eqref{Z1}, and expressing $\mathbf{x}$ in units of the block lengths $(l_{x},l_{y})$, we obtain
\begin{align}
  \!\!\!F&\equiv{\rm lim}^{\mathbf{N}_{b}}_{\rm s}\left[(N_{b})^{-1}\sum_{i=1}^{N^{x}_{b}}\sum_{j=1}^{N^{y}_{b}}{\rm f}(\varphi_{i\pm1j},\varphi_{ij},\varphi_{ij\pm1})\right]\nonumber\\
  &=\int_{A}\!{\rm d}\mathbf{x}\!\left[{\rm f}(\varphi(\mathbf{x}))+\frac{1}{2}\nabla\varphi(\mathbf{x})^T\boldsymbol{\kappa}(\varphi(\mathbf{x}))\nabla \varphi(\mathbf{x})\right]\!,
    \label{F}
\end{align}
where $A=[-L_{x}/2,L_{x}/2]\times[-L_{y}/2,L_{y}/2]$, and  the \emph{local free energy density} ${\rm f}(\varphi)$ and concentration-dependent \emph{gradient energy coefficient} $\boldsymbol{\kappa}(\varphi)$ are given by}
\begin{figure*}[ht!]
    \includegraphics[width = 0.9\textwidth]{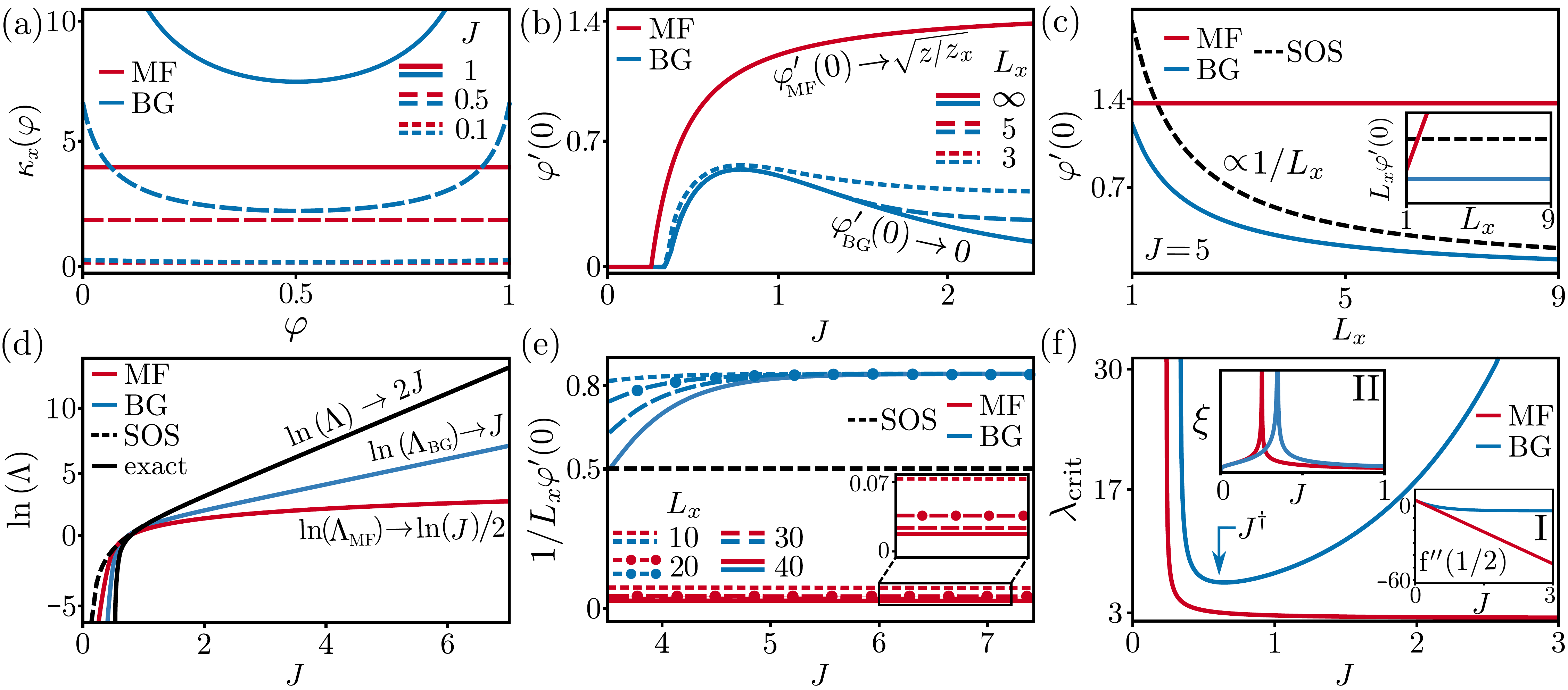}
    \caption{In all panels we consider a square lattice Ising strip with $\{z_{x},z_{y}\}=\{2,2\}$.\ Red, blue, and black solid/dashed lines correspond to MF, BG, and exact/SOS results, respectively.~(a) $x$-component of the gradient energy coefficient, $\kappa_{x}$, given by Eq.~\eqref{kappa} as a function of $\varphi$ for $J\in\{0.1,0.5,1\}$.\ (b)-(c) Interface steepness $\varphi^{\prime}(0)=\partial_x\varphi|_{x=0}$ of the equilibrium concentration profile as a function of $J$ for fixed $L_{x}\in\{3,5,\infty\}$ (b), and as a function of $L_{x}$ for fixed $J=5$ (c).\ Inset of (c):\ Rescaled steepness $L_{x}\varphi^{\prime}(0)$.\ (d) Interface stiffness $\Lambda$ defined in Eq.~\eqref{Gamma} as a function of $J$ on a logarithmic scale. The exact result (black line) is given by Eq.~\eqref{lambda}. \ (e) $1/L_{x}\varphi^{\prime}(0)$ as a function of $J$ for fixed
    $L_{x}\in\{10,20,30,40\}$. Blue lines converge to the value $\Delta_{\rm BG}\approx0.835$, corresponding to the instantaneous interface width (see also Eq.~\eqref{delta}). Inset: Blow-up of the MF result.\ (f) Critical stability wavelength $\lambda_{\rm crit}=2\pi[-\kappa_{x}(1/2)/{\rm f}''(1/2)]^{1/2}$ as a function of $J$; The blue arrow indicates $J^{\dagger}$ in Eq.~\eqref{Jlambda} where $\lambda^{\rm BG}_{\rm crit}$ attains a minimum.\ Inset I:\ Curvature of the free energy barrier ${\rm f}^{\prime\prime}(1/2)$. Inset II:\ Bulk correlation length $\xi$ defined in Eq.~\eqref{bulkcorrelationlength}.}
    \label{Fig4}
\end{figure*}
\blue{\begin{widetext}
\begin{equation}
    {\rm f}(\varphi)
    =2[z_{x}J_{x}\hat{\zeta}^{x}_{\dagger}{+}z_{y}J_{y}\hat{\zeta}^{y}_{\dagger}{-}1/4]+(1{-}z)[\Xi(\varphi)+\Xi(1-\varphi)]{+}\{(z_{x}/2)[\Xi(\varphi{-}\hat{\zeta}^{x}_{\dagger}){+}\Xi(1{-}\varphi{-}\hat{\zeta}^{x}_{\dagger}){+}2\Xi(\hat{\zeta}^{x}_{\dagger})]{+}(x\leftrightarrow y)\},
    \label{f_uniform_BG_per_site}
\end{equation}
\end{widetext}
\begin{equation}
    \boldsymbol{\kappa}(\varphi)=\frac{\mathbf{z}(\exp{4\mathbf{J}}-1)}{4(1+4(\exp{4\mathbf{J}}-1)\varphi(1-\varphi))^{1/2}},
    \label{kappa}
\end{equation}
with $\mathbf{J}={\rm diag}(J_{x},J_{y})$. Eqs.~\eqref{F}-\eqref{kappa} are the main result of the theoretical work presented here. 
The MF
analogs are obtained by taking the weak interaction limit
$\lim_{J_{x,y}\rightarrow0}{\rm f}(\varphi){=}{\rm f}_{\rm
  MF}(\varphi){+}\mathcal{O}(J_{x,y}^{2})$ where ${\rm f}_{\rm
  MF}(\varphi)$ is given in \cite{Note2}, and
$\lim_{J_{x,y}\rightarrow0}\boldsymbol{\kappa}(\varphi){=}\boldsymbol{\kappa}_{\rm
  MF}{+}\mathcal{O}(J_{x,y}^{2})$ with $\boldsymbol{\kappa}_{\rm
  MF}{=}\mathbf{z}\mathbf{J}$. Similarly, $\rm f_{\rm MF}(\varphi)$ can be obtained with the substitution $\hat{\zeta}^{x,y}_{\dagger}{\rightarrow}\hat{\zeta}^{x,y}$ in Eq.~\eqref{f_uniform_BG_per_site}. Note that $\boldsymbol{\kappa}_{\rm MF}$ is
independent of $\varphi$, in agreement with regular solution theory
\cite{cahn1958freeI}.\ 
Fig.~\ref{Fig4}a displays $\kappa_x$ as a function of
  $\varphi$ for BG and MF theory (blue and red
  lines, respectively). Here we observe a large \emph{entropic} penalty of
  inhomogeneities at $\varphi \to \{0,1\}$ (see \cite{Note2}) \emph{not}
  accounted for in MF theory.\\}
\section{Analysis of free energy functional accounting for pair correlations}\label{SecVI}
\subsection{Equilibrium interface profile}
In subsequent analysis we consider an isotropic interaction strength $J_{x}=J_{y}=J$.  The 
equilibrium profile minimizes $F$, i.e.\ it is the solution of
$\delta F/\delta \varphi(\mathbf{x})=0$.\ 
We now show that BG and MF theories
predict starkly different behavior for moderate and strong
interactions---MF theory fails to account for the interface broadening explained in Sec.~\ref{SecII}. First, considering Fig.~\ref{fig1}, we focus on the square lattice Ising strip ($L_{y}\gg L_{x}$) 
where the 
magnetization
varies only in the $x$ direction,
i.e.\ $\varphi(\bx)=\varphi(x)$. 
The
profile is obtained as the solution of a nonlinear second order
differential equation that we solve numerically. The boundary conditions are given by $\varphi(-L_{x}/2){=}\varphi_{\rm min}$ and $\varphi(L_{x}/2){=}1{-}\varphi_{\rm min}$,
where 
\blue{
\begin{equation}
    \varphi_{\rm min}\equiv\arg\inf_{0<\varphi\leq1/2}{\rm
  f}(\varphi)
\end{equation}}
is the
co-existing state determined by the location of the left
minimum of ${\rm f}(\varphi)$. Note that ${\rm f}(\varphi)$ is mirror symmetric around $\varphi=1/2$ in the absence of an external field.\ 
Above the critical coupling  $J{>} J_{\rm crit}$,
where \cite{blom2021criticality}
\blue{
\begin{equation*}
    J_{\rm BG,
  crit}\equiv\ln{(z/[z-2])}/2, \
  J_{\rm MF,
  crit}\equiv1/z,
\end{equation*}}
${\rm f}(\varphi)$ has two local minima resulting in a nonuniform
$\varphi(x)$. For $J{\le} J_{\rm crit}$ the profile is
uniform. We fix the ensemble interface location such that $\varphi(0)=1/2$ \footnote{Fixing the
    ensemble averaged interface position is not equal to fixing
    the position of instantaneous profiles. Thus, the interface
    location along individual trajectories may still fluctuate.}.\\ 
\indent Qualitative differences between the profiles predicted by BG and MF
theory are seen already in Fig.~\ref{fig1}c-d.\ 
In particular, BG concentration profiles depend
non-monotonically on $J$, which is confirmed by Monte-Carlo (MC)
simulations of the Ising model (for simulation details see \cite{Note2}), whereas MF interfaces become monotonically
steeper. By comparing with Fig.~\ref{fig1}b we observe that interface broadening correlates with interface delocalization. This
is further analyzed systematically in Fig.~\ref{Fig4}.\\
\indent First, we inspect in Fig.~\ref{Fig4}b the interface steepness $\varphi^{\prime}(0)$.
In stark contrast to MF theory predicting
a steepening interface independent of lattice size $L_{x}$, BG profiles are
non-monotonic in $J$ beyond a sufficient $L_{x}$
  due to interface delocalization.
  To verify that this is \emph{no} artifact, we compare our results with the
  \emph{solid-on-solid} (SOS) model for the square lattice Ising strip
  ($z=4$), which becomes exact in the limit
  $J\rightarrow\infty$, and is known to include interface
  delocalization \cite{ciach1987scaling,PhysRevB.34.1932,albano2000monte,PhysRevB.49.1092,PhysRevB.47.7519}.\ The
  SOS model yields \cite{albano2000monte,PhysRevB.49.1092,PhysRevB.47.7519}
  \begin{equation}
  \lim_{J\rightarrow\infty}\varphi_{\rm SOS}(x)=1/2{+}x/L_x{+}\sin(2\pi x/L_{x})/2\pi,
  \label{phiSOS}
  \end{equation}
  hence
  $\lim_{J\rightarrow\infty}\varphi^{\prime}_{\rm SOS}(0)=2/L_{x}$. In
  Fig.~\ref{Fig4}c we show the interface steepness as a function of
  $L_{x}$ for fixed $J$, and find that the SOS and BG results display
  the same scaling w.r.t.\ $L_{x}$ (see Fig.~\ref{Fig4}c inset), whereas the MF result
  is in fact independent of $L_{x}$.
  
  Further verification is given by
  the interface stiffness, which is the free energy difference between the non-uniform equilibrium profile $\varphi(x)$ and a uniform equilibrium profile $\varphi_{\rm min}$, and reads (see Eq.~(2.15) in \cite{cahn1958freeI})
  \begin{equation}
    \Lambda\equiv2\int_{\varphi_{\rm
      min}}^{1-\varphi_{\rm min}}[\kappa(\varphi)({\rm
      f}(\varphi)-{\rm f}(\varphi_{\rm
      min}))]^{1/2}d\varphi,
      \label{Gamma}
  \end{equation}  
which is depicted in
Fig.~\ref{Fig4}d. Note that surface tension is related to surface stiffness via $\sigma=\arcsinh{(\Lambda)}$ \cite{PhysRevB.47.7519}.\  
 The exact result is given by Eq.~\eqref{lambda}, while the SOS model yields
 $\Lambda_{\rm SOS}=\cosh{(2J)}-1$
 \cite{PhysRevB.47.7519}, and converges to the
exact result for large $J$,
i.e.\ $\lim_{J\rightarrow\infty}\ln{(\Lambda_{\rm
    SOS})}\simeq 2J$.\ Notably, the BG result is 
considerably more accurate than the MF prediction (compare blue and red line with the black line in Fig.~\ref{Fig4}d), and also displays a correct exponential scaling,
$\lim_{J\rightarrow\infty}\ln{(\Lambda_{\rm
    BG})}\simeq J$, in stark contrast to the square-root MF
scaling, $\lim_{J\rightarrow\infty}\ln{(\Lambda_{\rm
    MF})}\simeq\ln{(\sqrt{J})}$.
\subsection{Disentangling interface
  delocalization}\label{secdisentangle}
By exploiting the mapping of instantaneous interface positions onto a Brownian excurison problem (see Sec.~\ref{SecIV}) we can disentangle
  interface delocalization from the instantaneous interface
  width, $\Delta$, in the large $J$ limit where the
  instantaneous interface positions become
    asymptotically uniformly distributed, i.e.~
    \blue{\begin{equation}
    \lim_{J\rightarrow\infty}p_{\rm int}(x;J)=L^{-1}_{x}\mathbbm{1}_{|x|< L_{x}/2},
    \end{equation}}
    with $\mathbbm{1}_{|x|<L_{x}/2}$ equal to $1$ when $|x|{<}L_{x}/2$ and $0$ otherwise (see derivation
  in Sec.~\ref{SecIVB}). Let us assume that for $J\gg 1$ each instantaneous profile $\varphi_{\rm i}(x)$ is given by
  some continuous function $f(x/\Delta+b_{j}):\mathbbm{R}\to [0,1]$
  obeying $\lim_{x\rightarrow\pm\infty}f(x)=(1\pm1)/2$, where $\Delta>0$
  and $b_j$ describe the width and position of the $j$th instantaneous
  interface.\ The ensemble averaged profile is then given by the convolution of $f(x)$ with the probability density to have a certain shift $b$, i.e.\
  \blue{\begin{equation}
  \lim_{J\rightarrow\infty}\varphi(x)=L^{-1}_{x}\int
  f(x/\Delta+b)\mathbbm{1}_{|b|<L_{x}/2}db.
  \end{equation}}
  We can now compute the interface steepness 
  and find
  ${\lim_{J\rightarrow\infty}L_{x}\varphi^{\prime}(0){=}\Delta^{-1}(f(L_{x}/2){-}f(-L_{x}/2))}$.\ Finally,
  taking the large-$L_{x}$ limit, we obtain
  \begin{equation}
    \lim_{L_{x}\rightarrow\infty}\lim_{J\rightarrow\infty}1/L_{x}\varphi^{\prime}(0)=\Delta,
    \label{delta}
  \end{equation}
  which thereby disentangles interface delocalization from the
 instantaneous interface width $\Delta$.\ For 
  the Ising strip
this yields $\Delta_{\rm SOS}=0.5$ within the SOS model obtained from Eq.~\eqref{phiSOS}, and $\Delta_{\rm
  BG}\approx 0.835$ with the BG approximation (see
Fig.~\ref{Fig4}e).\ Hence, while the ensemble averaged steepness
vanishes in the large coupling limit due to interface delocalization,
instantaneous realizations maintain a nonzero interface steepness with
uniformly distributed interface positions.\ Importantly, MF theory does \emph{not}
account for delocalization-induced interface broadening and therefore
predicts $\Delta_{\rm MF}\rightarrow 0$ (see inset of Fig.~\ref{Fig4}e).
\subsection{Spinodal decomposition}
Having established the
  physical consistency of Eqs.~(\ref{F}-\ref{kappa}), we now address phase
separation, and determine the length scales on
which inhomogeneities are stable by performing a linear
stability analysis on the total free energy density around a uniform
concentration profile, i.e.\ $\varphi(x){=}\varphi_{0}{+}a\sin{(qx)}$ with $|a|{\ll}\min{(\varphi_{0},1-\varphi_{0})}$ (the symmetry of the problem imposes odd inhomogeneities). Stable
perturbations lower the total free energy density, $\Delta F\equiv F[\varphi(x)]-F[\varphi_{0}]\leq0$,
yielding an upper bound on stable wavevectors $q\leq q_{\rm crit}$ with the critical wavevector given by  (see also \cite{Note2})
\blue{\begin{align}
    q_{\rm crit}&\equiv \sqrt{-{\rm f}^{\prime\prime}(\varphi_{0})/\kappa_{x}(\varphi_{0})}\nonumber\\
    &=\sqrt{\frac{2(z-2)(1+4(\e{4J}{-}1)\varphi_{0}(1{-}\varphi_{0}))^{\frac{1}{2}}-2z}{z_{x}\varphi_{0}(1-\varphi_{0})(\e{4J}-1)}},
    \label{qc-BG}
\end{align}}
where ${\rm
  f}^{\prime\prime}(\varphi_{0})={\rm d}^{2}{\rm f}(\varphi)/{\rm d}\varphi^{2}|_{\varphi=\varphi_{0}}$
is the curvature of the free energy barrier. The critical wavevector translates into
a critical wavelength $\lambda_{\rm crit}=2\pi/q_{\rm crit}$, above
which perturbations are stable. Fig.~\ref{Fig4}f depicts $\lambda_{\rm
  crit}$ for a square lattice with $\varphi_{0}=1/2$ as a function of $J$. Similar to the results shown in 
Fig.~\ref{Fig4}b, $\lambda_{\rm crit}$ displays a non-monotonic
trend within BG theory (blue lines) \footnote{The non-monotonic dependence of $\lambda^{\rm BG}_{\rm crit}$ on $J$ in fact persist for any background
concentration $0<\varphi_{0}<1$; see \cite{Note2}.} that is contrasted by a monotonic
attenuation in the MF theory (red lines). The interaction strength
minimizing  $\lambda_{\rm crit}$ in the BG theory, i.e.\ the interaction strength $J$
allowing for the widest range of stable wavelengths can be determined exactly and
reads (see blue arrow in
Fig.~\ref{Fig4}f)
\blue{\begin{equation}
    J^{\dagger}(\varphi_{0})=\frac{1}{4}\ln{\left(1+\frac{z(2+\sqrt{z-1})-2}{(z-2)^{2}\varphi_{0}(1-\varphi_{0})}\right)},
    \label{Jlambda}
\end{equation}}
with the corresponding $\lambda_{\rm
  crit}(J^{\dagger})$ given in \footnote{$\lambda_{\rm crit}^{\rm BG}(J^{\dagger})$ is independent of the uniform background
concentration $0<\varphi_{0}<1$; see
\cite{Note2}.}. 
The non-monotonicity of
$\lambda_{\rm crit}$ can be understood by inspecting 
how the curvature of
the barrier depends on $J$.\ 
In particular,
the BG curvature
converges, 
$\lim_{J\rightarrow\infty}{\rm
  f}^{\prime\prime}(1/2){=}2(2{-}z)$ (see Fig.~\ref{Fig4}f, inset I), whereas the free energy penalty of
inhomogeneities increases exponentially, eventually increasing
$\lambda_{\rm crit}$.\  MF
overestimates the curvature of the barrier, and underestimates the free energy penalty of inhomogeneities, leading a decrease in $\lambda^{\rm MF}_{\rm
  crit}$.\ Notably, the bulk correlation length \cite{parry2019goldstone}
\blue{\begin{equation}
    \xi \equiv\sqrt{\kappa_{x}(\varphi_{\rm
    min})/{\rm f}^{\prime\prime}(\varphi_{\rm
    min})},
    \label{bulkcorrelationlength}
\end{equation}}
displays qualitatively the
same behavior in both theories (see Fig.~\ref{Fig4}f, inset II), since 
the MF free energy density is relatively accurate near
the local minimum $\varphi_{\rm min}$, but inaccurate near the
barrier
(see \cite{Note2}
and \cite{blom2021criticality}). 
\subsection{Implications for nucleation}
\begin{figure}[t!]
    \includegraphics[width = 0.47\textwidth]{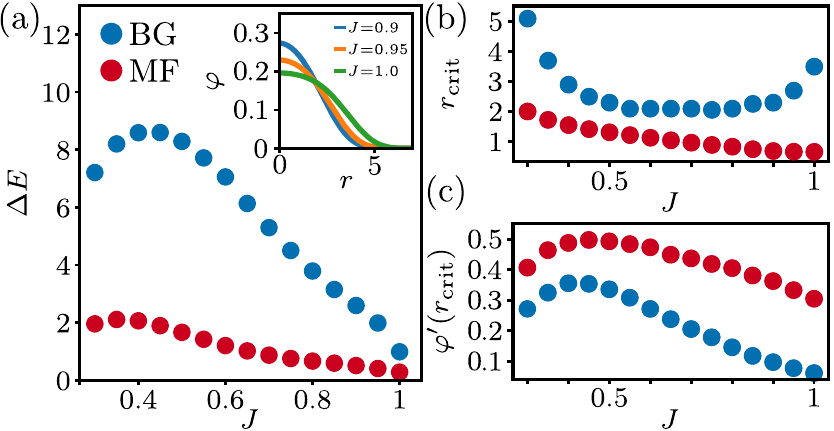}
    \caption{Numerical simulations of critical nuclei of the radially symmetric Cahn-Hilliard equation with the BG (blue) and MF (red) free energy for a hexagonal coordination $\{z,z_{x}\}=\{6,4\}$).
(a) Free energy difference $\Delta E$ between the critical nucleus and the homogeneous state as a function of the interaction strength $ J$.
The inset shows critical profiles $\varphi(r)$ for three values of $ J$.
(b) Radius~$r_\mathrm{crit}$ and (c) interface steepness $\varphi'(r_\mathrm{crit})$ of the critical nucleus as a function of $ J$.}
    \label{Fig5}
\end{figure}

We next investigate, in
Fig.~\ref{Fig5}, how interface broadening affects
nucleation, by determining
minimal free energy paths (the
  reaction coordinate and 
  method are described in Sec.~IX of \cite{Note2}). 
The inset in Fig.~\ref{Fig5}a suggests that critical nuclei become
less dense and wider as $J$ becomes larger.
Indeed, we find that correlations captured by BG theory lead to larger critical nuclei (Fig.~\ref{Fig5}b), shallower interfaces
(Fig.~\ref{Fig5}c), and that the increasing trend with $ J$ is
only captured by BG theory, which is reminiscent of the results shown
in Fig.~\ref{Fig4}. 
Most importantly, BG theory predicts that the nucleation barrier
$\Delta E$ is approximately four times larger than predicted by MF
(Fig.~\ref{Fig5}a), implying a strong reduction of nucleation
rates~\cite{Kramers,Langer,Hanggi1990,Duality}. 

To understand why interface delocalization affects
  nucleation we note that shifting the interface position
  corresponds to a growing/shrinking nucleus which alters the
  free energy.\ Instantaneous interfaces are
  still affected by interface translation and capillary-wave fluctuations.\ However, in contrast to the strip, distinct instantaneous
  interface configurations are 
  \emph{not} iso-energetic.\ The weighting
  by the respective free energy of the configuration ultimately gives rise to broadening,
  and thus larger critical nuclei and higher nucleation barriers.
\section{Conclusion}\label{SecVII}
By directly computing the thermodynamic limit of a spatially inhomogeneous Ising model on 
general lattices
within the Bethe-Guggenheim approximation we
derived a Cahn-Hilliard\blue{-type phase-field} free energy that accounts for nearest-neighbor
pair-correlations.~Strong interactions were shown to give rise to a
delocalization-induced interface 
  broadening confirmed by exact results for the two-dimensional Ising model, 
a strong reduction of nucleation kinetics due to an amplification of the free energy barrier to
  nucleation, and a non-monotonic dependence of critical
  nucleus size on interaction strength.
  These effects are the result of an entropy-driven interplay between
    capillary-wave and interface-position fluctuations at sufficiently
strong coupling, and pair
correlations are required to correctly
  account for them.\ Pair
correlations enforce a thermodynamically optimal configuration of
defects, and are thus an essential determinant of interfaces and
condensates in the strong interaction limit that so far have been
overlooked. \blue{By neglecting correlations, mean-field reasoning
  inherently disregards correlations and thus cannot account for local defects and their entropic
  stabilization, and is thus thermodynamically inconsistent in the
  intermediate- and strong-interaction regime.} Our results allow for
generalizations to three dimensions, more than two constituents,
\blue{and conservation laws,} which will be addressed in forthcoming publications.\vspace{0.2cm}\\

\textit{Acknowledgments.---}The financial support from the German
Research Foundation (DFG) through the Emmy Noether Program GO 2762/1-2
(to AG) and Project ZW 222/3-1 (to DZ), and the Max Planck Society (to
DZ and KB), in case of KB in  the form of an IMPRS fellowship are gratefully acknowledged.
\bibliographystyle{apsrev4-1.bst}
\bibliography{Broadening.bib}

\let\oldaddcontentsline\addcontentsline
\renewcommand{\addcontentsline}[3]{}
\let\addcontentsline\oldaddcontentsline

\clearpage
\newpage
\onecolumngrid
\renewcommand{\thefigure}{S\arabic{figure}}
\renewcommand{\theequation}{S\arabic{equation}}
\setcounter{equation}{0}
\setcounter{figure}{0}
\setcounter{page}{1}
\setcounter{section}{0}
\begin{center}\textbf{Supplementary Material for:\\Thermodynamically Consistent Phase-Field Theory Including Nearest-Neighbor Pair
  Correlations Explains Failure of Mean-Field Reasoning}\\[0.2cm]
Kristian Blom, Noah Ziethen, David Zwicker, and Alja\v{z} Godec\\
\emph{Mathematical bioPhysics Group, Max Planck Institute for Multidisciplinary Sciences, Am Fa\ss berg 11, 37077 G\"ottingen}\\[0.6cm]\end{center}

\date{\today}

\maketitle
\noindent The sections in this Supplementary Material (SM) are
  organized in the order they appear in the main article. First, we
  present in Sec.~\ref{secI}     a detailed description of Monte-Carlo
  simulations that are shown in Fig.~1 in the main article. Sections~\ref{secIII}-\ref{secV} are devoted to the derivation of the Cahn-Hilliard theory starting from a two-dimensional Ising model using the mean field (Sec.~\ref{secIV}) and Bethe-Guggenheim (Sec.~\ref{secV})
approximations, respectively. The latter is also discussed in Sec.~V in the main article, but here we provide some more technical details. Next, in Sections~\ref{secVI}-\ref{secVII} we analyze the field theories by determining the one-dimensional 
equilibrium concentration profile, interfacial steepness, interfacial
width, and critical wavelength of stable perturbations. In Sec.~\ref{secVIII} we probe the accuracy of both approximations by comparing them with exact results for system sizes which are amendable to exact solutions. Finally, in Sec.~\ref{secIX} we present details
on the numerical simulations of nucleation shown in Fig.~5 in the main article.

\tableofcontents
\section{Monte-Carlo simulations of the Ising model}\label{secI}
\noindent Here we provide details on the Monte-Carlo (MC)
  simulations which we used to determine the ensemble averaged concentration
  profile and histograms of instantaneous interface locations
  displayed in Fig.~1 in the main article.
\subsection{Lattice setup and initial configuration} 
\noindent We performed MC simulations of the nearest-neighbor
interacting ferromagnetic Ising model on the square
lattice of size $(N^{x}_{\sigma}=40)\times
(N^{y}_{\sigma}\in\{80,90,100,110,120,130\})$ with single spin-flip 
dynamics in the bulk and two-spin-exchange dynamics at the
boundary columns located at $i=\pm
N^{x}_{\sigma}/2$. We considered various values of
  $N^{y}_{\sigma}$ to benchmark our simulations against known
  theoretical predictions (see Sec.~\ref{secIE}). We imposed periodic boundary conditions
in the vertical direction (i.e.\ along the columns) and free boundary
conditions in the horizontal direction (i.e.\ along the rows), whereby
we constrained the total magnetization on the left/right boundary (see
below). Let
$N^{\downarrow}_{i}$ with
$i\in\{-N^{x}_{\sigma}/2,...,N^{x}_{\sigma}/2\}$ denote the number of
down spins in column $i$. To induce a non-uniform concentration
profile, and in anticipation of known exact results for the bulk
concentration values \cite{sPhysRev.85.808}, we fixed the number of down
spins at the boundaries to be 
\begin{equation}
    N^{\downarrow}_{\pm N^{x}_{\sigma}/2}=\frac{N^{y}_{\sigma}}{2}\left(1\pm{\rm Re}([1-\sinh^{-4}{(2J)}]^{1/8})\right),
    \label{b.c.MC}
\end{equation}
where $J$ is the coupling strength in units of $k_{\rm B}T$. Spins located at the boundaries can exchange \emph{only} within the same
column, and therefore the total number of up/down spins at the
boundaries is conserved throughout the simulation. Spins in the bulk
are initially prepared in a high-coupling
  configuration (i.e.\ aligned) with a vertical interface placed at some random horizontal location in the lattice. Starting from a high-coupling configuration has the advantage that the simulations do not get stuck in frozen sub-optimal states where multiple interfaces are created \cite{sPhysRevE.63.036118,sPhysRevE.80.040101}. 
\subsection{Acceptance rate}
\noindent For single spin-flip dynamics let $\{\sigma_{j}\}^{\prime}_{i}$ denote the spin configuration obtained by flipping spin $i$ while keeping
the configuration of all other spins fixed,
i.e.,\ $\{\sigma_{j}\}^{\prime}_{i}\equiv(-\sigma_{i},\{\sigma_{j\neq
  i}\})$. Moreover, let $p_{i}(\{\sigma_{j}\})$ denote the acceptance
rate from $\{\sigma_{j}\}$ to $\{\sigma_{j}\}^{\prime}_{i}$ and
$\Delta\mathcal{H}_{i}(\{\sigma_{j}\})\equiv\mathcal{H}(\{\sigma_{j}\}^{\prime}_{i})-\mathcal{H}(\{\sigma_{j}\})$
the energy difference (in units of $k_{\rm B}T$) associated with the
transition. Using the Metropolis algorithm the acceptance rate for the single spin-flip takes the form \cite{sdoi:10.1063/1.1699114}
\begin{equation}
    p_{i}(\{\sigma_{j}\})=\min(1,\e{-\Delta \mathcal{H}_{i}(\{\sigma_{j}\})}).
\end{equation}
For two-spin-exchange dynamics let $\{\sigma_{j}\}^{\prime}_{ik}$ denote the spin
configuration upon interchanging the spins $\sigma_{i}$ and $\sigma_{k}$ while keeping the configuration
of all other spins fixed, i.e.,\ $\{\sigma_{j}\}^{\prime}_{ik}\equiv(\sigma_{i}\leftrightarrow\sigma_{k},\{\sigma_{j\neq (i,k)}\})$.
We denote with $p_{ik}(\{\sigma_{j}\})$ the acceptance rate from $\{\sigma_{j}\}$ to $\{\sigma_{j}\}^{\prime}_{ik}$ and $\Delta\mathcal{H}_{ik}(\{\sigma_{j}\})\equiv\mathcal{H}(\{\sigma_{j}\}^{\prime}_{ik})-\mathcal{H}(\{\sigma_{j}\})$ denotes the energy difference associated with the transition. Using the Metropolis algorithm the two-spin-exchange acceptance rate reads
\begin{equation}
    p_{ik}(\{\sigma_{j}\})=\min(1,\e{-\Delta \mathcal{H}_{ik}(\{\sigma_{j}\})}).
\end{equation}
\subsection{Simulation parameters}
\noindent For each value of the coupling strength $J$ and
vertical length $N^{y}_{\sigma}\in\{80,90,100,110,120,130\}$ we
performed $N_{\rm MC}=10^{5}$ MC simulations, where each individual
run contained $5\times 10^{8}$ MC steps. At each
  $1.9\times10^{7}$th MC step we took a snapshot of the configuration
  and stored the total energy, resulting in 26 (including the initial
  configuration) snapshots for each simulation run.
\subsection{Equilibration test: Energy fluctuations per spin}
\begin{figure}
    \centering
    \vspace*{-1cm}\hspace*{-0.4cm}\includegraphics[scale=0.5]{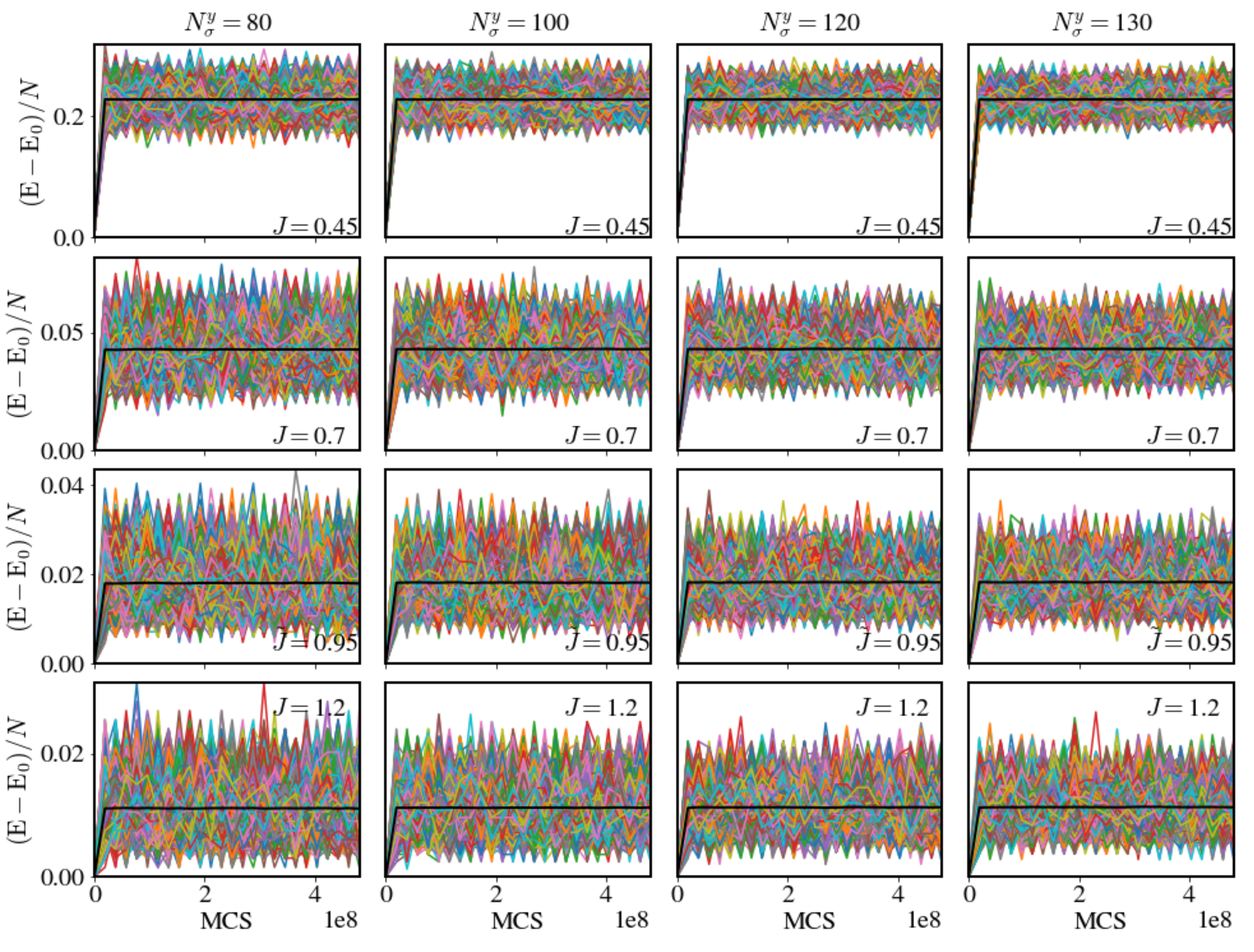}
    \caption{\textbf{Equilibration test:} Energy
        fluctuations per spin as a function of consecutively stored
        Monte-Carlo (MC) configurations (see text). In each plot we display the energy fluctuations per spin $({\rm E}-{\rm E}_{0})/N$ where ${\rm E}_{0}$ is the ground state energy conditioned on anti-symmetric boundary conditions and $N=N^{x}_{\sigma}N^{y}_{\sigma}$ with $N^{x}_{\sigma}=40$ for a subset of $10^{4}$ MC simulations (colored lines). The black solid line indicates the ensemble average energy fluctuation per spin. Plots in the same column have equal $N^{y}_{\sigma}\in\{80,100,120,130\}$, and plots in the same row have equal $J\in\{0.45,0.6,0.95,1.2\}$.}
    \label{fig-1SM}
\end{figure}
\noindent To assess whether the MC simulations
  reached equilibrium we analyzed the energy fluctuations per spin, and
  their corresponding ensemble average. In Fig.~\ref{fig-1SM} we display
  the energy fluctuations per spin for a subset of $10^{4}$
  simulations as a function of the MC steps (MCS) for various
  $J\in\{0.45,0.7,0.95,1.2\}$ and
  $N^{y}_{\sigma}\in\{80,100,120,130\}$. In each plot we observe that
  immediately after the initial snapshot the energy is fluctuating
  around an average steady state denoted with the black solid line,
  providing a first indication that the simulations have reached
  equilibrium (already at the first stored configuration). Note that
  in each plot all energies are initially increasing from zero since
  we subtract the ground state energy \emph{and} initialize the
  system in a high-coupling configuration which is identical to the
  ground state.  
\subsection{Benchmark test: Interfacial width and roughening}\label{secIE}
\begin{figure}
    \centering
    \includegraphics[width=0.95\textwidth]{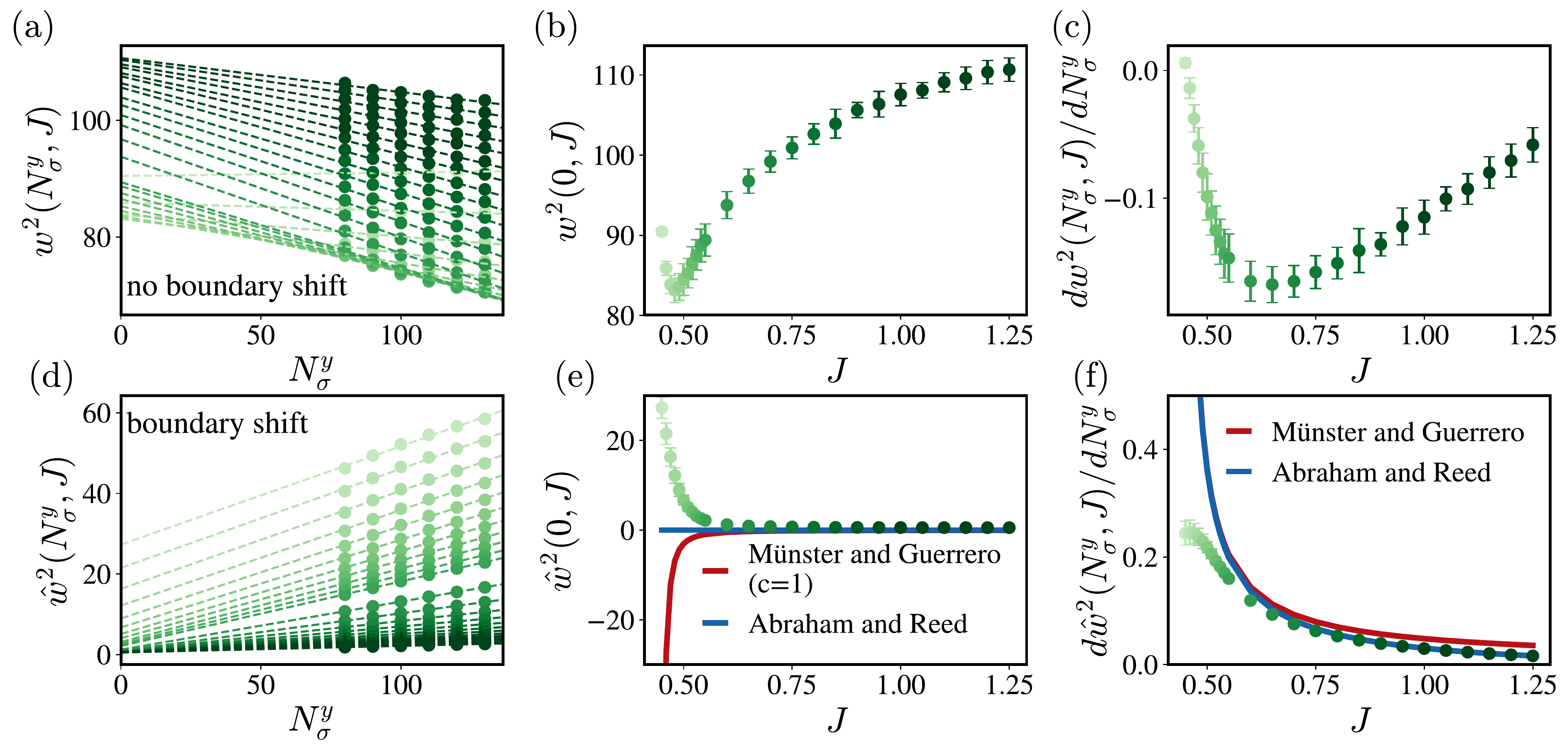}
    \caption{\textbf{Benchmark test:} Results in the
        bottom and top row are derived with and without applying the
        boundary-shift method, respectively. (a)-(d) Scaling of the
        interfacial width (a) $w^{2}(N^{y}_{\sigma},J)$ (no
        boundary shift) and (d)
        $\hat{w}^{2}(N^{y}_{\sigma},J)$ (boundary shift)
        w.r.t.~the vertical number of spins $N^{y}_{\sigma}$. Each
        point is obtained by averaging over $2.5\times10^{6}$
        equilibrated configurations. Dashed lines are obtained by
        weighted linear regression. Colors from light green to dark
        green correspond to increasing coupling strength
        $J$. (b)-(e) Intersection point of the interfacial
        width at $N^{y}_{\sigma}=0$ as a function of $J$. The
        standard deviation of each point is estimated with the
        Jackknife method. In (e) the red and blue lines are the
        theoretical predictions for the intersection point given in
        \cite{smunster2021interface, sPhysRevLett.47.545},
        respectively. (c)-(f) Slope of the interfacial width
        w.r.t.~$N^{y}_{\sigma}$ as a function of $J$. The
        standard deviation of each point is estimated with the Jackknife method.  In (f) the red and blue lines are the theoretical predictions for the slope given in \cite{smunster2021interface, sPhysRevLett.47.545}, respectively.}
    \label{fig0SM}
\end{figure}
\noindent To benchmark the performance of our MC
  simulations we computed the interfacial width
  $w^{2}(N^{y}_{\sigma},J)$ and compared our results with
  known theoretical results reported in \cite{sPhysRevLett.47.545,
    smunster2021interface}. The results from \cite{sPhysRevLett.47.545}
  predict $w^{2}(N^{y}_{\sigma},J)\propto
  N^{y}_{\sigma}/\sinh{(\sigma)}$ with
  $\sigma=2J+\ln{\tanh{J}}$. Analogously, the results
  from \cite{smunster2021interface} predict
  $w^{2}(N^{y}_{\sigma},J)=N^{y}_{\sigma}/12\sigma-c/2\pi\sigma^{2}$
  with $c\approx1$. Below we explain in detail how we determined the
  interfacial width and how it compares to the theoretical
  predictions. The resulting outcomes are shown in Fig.~\ref{fig0SM} and
  the comparison with the theoretical results are shown in
  Fig.~\ref{fig0SM}e-f.
\subsubsection{Ensemble averaged concentration profile and the boundary-shift method}
\noindent To compare our results with
  \cite{sPhysRevLett.47.545, smunster2021interface} we need to apply the
  so-called boundary shift method  \cite{smuller2005profile}
  where we shift the interface position of each instantaneous
  concentration profile to the center of the lattice. As a scientific
  exercise we also consider the resulting outcomes without applying
  the boundary shift method, for which the results are depicted in the
  top row of Fig.~\ref{fig0SM}. Let $\hat{\varphi}_{i,k}$ be the
equilibrated and boundary shifted concentration of
down spins in column $i$
of the $k$th MC simulation run. The ensemble average boundary-shifted concentration profile is given by
\begin{equation}
    \langle \hat{\varphi}_{i} \rangle=\frac{1}{N_{\rm MC}}\sum_{k=1}^{N_{\rm MC}}\hat{\varphi}_{i,k}.
    \label{vphi-jack}
\end{equation}
From Eq.~\eqref{vphi-jack} we can approximate the mean interfacial width
using the central difference method as follows
\begin{equation}
    \hat{w}^{2}(N^{y}_{\sigma},J)=\frac{\sum^{N^{x}_{\sigma}/2-1}_{i=-N^{x}_{\sigma}/2+1}i^2(\langle\hat{\varphi}_{i+1}\rangle-\langle\hat{\varphi}_{i-1}\rangle)}{\sum^{N^{x}_{\sigma}/2-1}_{i=-N^{x}_{\sigma}/2+1}\langle\hat{\varphi}_{i+1}\rangle-\langle\hat{\varphi}_{i-1}\rangle}-\left(\frac{\sum^{N^{x}_{\sigma}/2-1}_{i=-N^{x}_{\sigma}/2+1}i(\langle\hat{\varphi}_{i+1}\rangle-\langle\hat{\varphi}_{i-1}\rangle)}{\sum^{N^{x}_{\sigma}/2-1}_{i=-N^{x}_{\sigma}/2+1}\langle\hat{\varphi}_{i+1}\rangle-\langle\hat{\varphi}_{i-1}\rangle}\right)^{2}. 
    \label{w21}
\end{equation}
A similar definition holds for the interfacial width without applying the boundary-shift method, which we denote as $w^{2}(N^{y}_{\sigma},J)$. In Fig.~2a-d we plot $
  w^{2}(N^{y}_{\sigma},J)$ and $
  \hat{w}^{2}(N^{y}_{\sigma},J)$ with the green dots as a
  function of $N^{y}_{\sigma}$. Both results show a clear linear trend
  with $N^{y}_{\sigma}$, providing a first validation of the MC simulations. To obtain the variance of $\hat{w}^{2}(N^{y}_{\sigma},J)$ and $w^{2}(N^{y}_{\sigma},J)$ -- which we will use in the next section -- we used the Jackknife method which is explained below.
\subsubsection{Interfacial width and weighted linear regression}
\noindent To compare our results with those reported in \cite{sPhysRevLett.47.545, smunster2021interface} we need to extract the interception point $\hat{w}^{2}(J,0)$ and slope $d\hat{w}^{2}(J,N^{y}_{\sigma})/dN^{y}_{\sigma}$. To obtain both quantities we use weighted linear regression in combination with the Jackknife method. First we determine $\hat{w}^{2}(0,J)$ and slope $d\hat{w}^{2}(N^{y}_{\sigma},J)/dN^{y}_{\sigma}$ for fixed $J$ while removing one point from the data pool, which gives
\begin{equation}
    \{\hat{w}^{2}_{ j}(0,J),\frac{d\hat{w}^{2}_{ j}(N^{y}_{\sigma},J)}{dN^{y}_{\sigma}}\}=\min\limits_{(\alpha,\beta)}\sum_{\substack{N^{y}_{\sigma}=\{80,...,130\}\\N^{y}_{\sigma}\neq 70+10\times j}}\frac{(\alpha+\beta N^{y}_{\sigma}- \hat{w}^{2}(N^{y}_{\sigma},J))^{2}}{{\rm var}(\hat{w}^{2}(N^{y}_{\sigma},J))},
\end{equation}
where $j=\{1,...,6\}$. A similar definition holds for the intersection point and slope without applying the boundary shift method, which we denote as $w^{2}_{j}(0,J)$ and $dw^{2}_{j}(N^{y}_{\sigma},J)/dN^{y}_{\sigma}$, respectively. Finally the Jackknife ensemble averages and variances are given by
\begin{eqnarray}
    \hat{w}^{2}(0,J) &=&\frac{1}{6}\sum_{j=1}^{6}\hat{w}^{2}_{j}(0,J), \ \ \ \ \ {\rm var}(\hat{w}^{2}(0,J))=\frac{5}{6}\sum_{j=1}^{6}(\hat{w}^{2}_{j}(0,J)-\hat{w}^{2}(0,J))^{2}, \nonumber \\
    \frac{d\hat{w}^{2}(N^{y}_{\sigma},J)}{dN^{y}_{\sigma}} &=&\frac{1}{6}\sum_{j=1}^{6}\frac{d\hat{w}^{2}_{ j}(N^{y}_{\sigma},J)}{dN^{y}_{\sigma}}, \
    {\rm var}\left(\frac{d\hat{w}^{2}(N^{y}_{\sigma},J)}{dN^{y}_{\sigma}}\right)=\frac{5}{6}\sum_{j=1}^{6}\left(\frac{d\hat{w}^{2}_{ j}(N^{y}_{\sigma},J)}{dN^{y}_{\sigma}}-\frac{d\hat{w}^{2}(N^{y}_{\sigma},J)}{dN^{y}_{\sigma}}\right)^{2}.
\end{eqnarray}
In Fig.~\ref{fig0SM}e-f we plot $\hat{w}^{2}(0,J)$ and
$d\hat{w}^{2}(N^{y}_{\sigma},J)/dN^{y}_{\sigma}$ together with
the standard deviation as a function of $J$. The theoretical
results given by \cite{sPhysRevLett.47.545, smunster2021interface} are
shown with the red and blue lines, respectively. For
$J\geq0.6$ we find a very good agreement between MC
simulations and theoretical predictions.  Notably, for the slope in
Fig.~\ref{fig0SM}f we find a remarkable agreement with the results of
\cite{sPhysRevLett.47.545}. For $J<0.6$ we approach the
critical coupling $J_{\rm crit}\approx0.441$, where the MC
results agree less well with theoretical     predictions due to
finite-size effects. This is expected since the correlation length
diverges around the critical coupling. 
\section{Coarse-grained partition function (a recap)}\label{secIII}
In the next two sections we derive a Cahn-Hilliard free energy based on the mean field (MF; Sec.~\ref{secIV}) and (BG; Sec.~\ref{secV}) approximation. To that aim let us recall the coarse-grained partition function per spin block $\mathcal{Z}_{ij}$, given by Eq.~(17) in the main article, which reads (for further details about the derivation of $\mathcal{Z}_{ij}$ see Sec.~IV in the main article)
\begin{equation}
    \mathcal{Z}_{ij}{=}\sum_{\varphi_{ij}}\smashoperator{\sum_{\zeta^{x,y}_{ij}}}\smashoperator{\sum_{\xi^{x,y\pm}_{ij}}}\Psi(\varphi_{ij},\zeta^{x,y}_{ij},\xi^{x,y\pm}_{ij})
    \e{{-}N^{b}_{\sigma}[\mathcal{H}_{\rm inter}+\mathcal{H}_{\rm intra}-\mathcal{C}]},
    \label{Z2}
\end{equation}
where $\Psi(\varphi_{ij},\zeta^{x,y}_{ij},\xi^{x,y\pm}_{ij})$ is the degeneracy of states, and the Hamiltonian is split into inter- and intrablock contributions
\begin{align}
    \mathcal{H}_{\rm inter}(\{\mathbf{b}_{i\pm1j},\mathbf{b}_{ij},\mathbf{b}_{ij\pm1}\})&=z_{x} J_{x}(\xi^{x+}_{ij}+\xi^{x-}_{ij})/2+(x\leftrightarrow y),
    \nonumber\\
    \mathcal{H}_{\rm intra}(\{\mathbf{b}_{ij}\})&=z_{x}J_{x}\zeta^{x}_{ij}+(x\leftrightarrow y),\nonumber\\
     \mathcal{C}&=(z_{x}J_{x}+z_{y}J_{y})/2,
    \label{Hinter2}
\end{align}
with $\xi^{x,y}_{ij}$ and $\zeta^{x,y\pm}_{ij}$ the fraction of intra- and interblock defects, defined in Eq.~(11) and (12) in the main article. Recall that the degeneracy of states obeys a normalization condition for $(J_{x},J_{y})=(0,0)$ given by
\begin{equation}
    \sum_{\zeta^{x,y}_{ij}}\sum_{\xi^{x,y\pm}_{ij}}\Psi(\varphi_{ij},\zeta^{x,y}_{ij},\xi^{x,y\pm}_{ij})\stackrel{!}{=}\binom{N^{b}_{\sigma}}{\varphi_{ij} N^{b}_{\sigma}},
    \label{norm2}
\end{equation}
where $N^{b}_{\sigma}$ is the number of spins inside a spin block (see Fig.~3 in the main article). Our aim is to evaluate Eq.~\eqref{Z2}. 
\section{Mean field approximation}\label{secIV}
\subsection{Approximation of the fraction of defects}
\noindent On the MF level we introduce the following approximation of the fraction of defects between two spin blocks $b_{ij}$ and $b_{mn}$:
\begin{equation}
    \hat{\zeta}_{\rm MF}(\varphi_{ij},\varphi_{mn})\equiv [\varphi_{ij}(1-\varphi_{mn})+\varphi_{mn}(1-\varphi_{ij})]/2.
    \label{Zeta_mf}
\end{equation} 
Thus, on the MF level we approximate the number of defects between
blocks $b_{ij}$ and $b_{mn}$ by the product of the \textit{spin down}
concentration in box $b_{ij}$ and \textit{spin up} concentration in
box $b_{mn}$, and vice versa. Making the substitutions
\begin{align*}
    \zeta^{x,y}_{ij}&\rightarrow\hat{\zeta}_{\rm
  MF}(\varphi_{ij},\varphi_{ij}),\\
\xi^{x\pm}_{ij}&\rightarrow\hat{\xi}^{x\pm}_{\rm MF}\equiv\hat{\zeta}_{\rm
  MF}(\varphi_{i\pm1j},\varphi_{ij}),\\
\xi^{y\pm}_{ij}&\rightarrow\hat{\xi}^{y\pm}_{\rm MF}\equiv\hat{\zeta}_{\rm
  MF}(\varphi_{ij\pm1},\varphi_{ij})
\end{align*}
we see that $\mathcal{H}_{\rm inter}$ and $\mathcal{H}_{\rm intra}$ 
inside the exponent of Eq.~\eqref{Z2} become independent of the
variables
$(\zeta^{x,y}_{ij},\xi^{x,y\pm}_{ij})$ and only depend on $\varphi_{ij}$. Therefore
we can directly use Eq.~\eqref{norm2} to perform the four inner sums in Eq.~\eqref{Z2}. This results in the MF partition function
\begin{equation}
    \mathcal{Z}_{ij}^{\rm MF}=\sum\limits_{\varphi_{ij}}\binom{N^{b}_{\sigma}}{\varphi_{ij} N^{b}_{\sigma}}\e{-N^{b}_{\sigma}[z_{x}J_{x}(\hat{\zeta}_{\rm MF}+\{\hat{\xi}^{x+}_{\rm MF}+\hat{\xi}^{x-}_{\rm MF}\}/2)+(x\leftrightarrow y)-\mathcal{C}]}.
    \label{Z_mf}
\end{equation}
To evaluate the sum over $\varphi_{ij}$ in Eq.~\eqref{Z_mf} we employ
the maximum term method, 
and take the maximum term of the continuous summand in the
thermodynamic limit, which is defined in Eq.~(9) in the main article. 
To that end we introduce the MF free energy density
\begin{align}
    {\rm f}_{{\rm MF}}(\varphi_{i\pm1j},\varphi_{ij},\varphi_{ij\pm1})&\equiv{\rm lim}^{\mathbf{N}_{\sigma}}_{\rm s}\left[-(N^{b}_{\sigma})^{-1}\ln{\left(\binom{N^{b}_{\sigma}}{\varphi_{ij} N^{b}_{\sigma}}\e{-N^{b}_{\sigma}[z_{x}J_{x}(\hat{\zeta}_{\rm MF}+\{\hat{\xi}^{x+}_{\rm MF}+\hat{\xi}^{x-}_{\rm MF}\}/2)+(x\leftrightarrow y)-\mathcal{C}]}\right)}\right] \nonumber \\
    &=\Xi(\varphi_{ij})+\Xi(1-\varphi_{ij})+\{z_{x}J_{x}(\hat{\zeta}_{\rm MF}+\{\hat{\xi}^{x+}_{\rm MF}+\hat{\xi}^{x-}_{\rm MF}\}/2)+(x\leftrightarrow y)\}-\mathcal{C},
    \label{fj_mf}
\end{align}
where we used Stirling's approximation $\ln{(n!)}=
\Xi(n)-n+\mathcal{O}(\ln{(n)})$ with $\Xi(n)\equiv n\ln{(n)}$ to
evaluate the logarithm of the binomial coefficient. Note that so far
we have only taken the thermodynamic limit of the spins. This makes
$\varphi_{ij}\in[0,1]$ a continuous variable, as well as $\hat{\zeta}_{\rm
  MF}\in[0,1/4]$ and $\hat{\xi}^{x,y\pm}_{\rm
  MF}\in[0,1/2]$. Upon considering the
thermodynamic limit of the spin blocks, we can employ two different strategies (as proposed in the main article): 
\begin{enumerate}
    \item First optimize ${\rm f}_{{\rm MF}}(\varphi_{i\pm1j},\varphi_{ij},\varphi_{ij\pm1})$ over $\varphi_{ij}$ and finally apply ${\rm lim}^{\mathbf{N}_{b}}_{\rm s}\left[\cdot\right]$.
    \item First apply ${\rm lim}^{\mathbf{N}_{b}}_{\rm s}[{\rm f}_{{\rm MF}}(\varphi_{i\pm1j},\varphi_{ij},\varphi_{ij\pm1})]$ and then optimize the resulting free energy functional.
\end{enumerate}
Below we carry out both, and show that they give equivalent results for
the resulting concentration profile. Only the second strategy, however, leads to a Cahn-Hilliard type free energy functional. 
\subsection{Evaluating the partition function: Strategy 1}
\noindent Using the maximum term method we need to find the location $\varphi_{ij}$ which renders ${\rm f}_{{\rm MF}}(\varphi_{i\pm1j},\varphi_{ij},\varphi_{ij\pm1})$ minimal, yielding the equation
\begin{equation}
    \partial_{\varphi_{ij}}\left[{\rm f}_{{\rm MF}}(\varphi_{i\pm1j},\varphi_{ij},\varphi_{ij\pm1})+\sum_{k=\pm1}\left({\rm f}_{{\rm MF}}(\varphi_{i+k\pm1j},\varphi_{i+kj},\varphi_{i+kj\pm1})+{\rm f}_{{\rm MF}}(\varphi_{i\pm1j+k},\varphi_{ij+k},\varphi_{ij+k\pm1})\right)\right]\stackrel{!}{=}0,
    \label{Optim_MF}
\end{equation}
where $\partial_{\varphi_{ij}}\equiv\partial/\partial\varphi_{ij}$. Note that aditionally to ${\rm f}_{{\rm MF}}(\varphi_{i\pm1j},\varphi_{ij},\varphi_{ij\pm1})$, four extra terms enter Eq.~\eqref{Optim_MF} which also contain an explicit dependence on $\varphi_{ij}$. The solution to Eq.~\eqref{Optim_MF} can be cast into the following set of difference equations
\begin{equation}
    z_{x}J_{x}(\varphi_{i+1j}-2\varphi_{ij}+\varphi_{i-1j})+z_{y}J_{y}(\varphi_{ij+1}-2\varphi_{ij}+\varphi_{ij-1})=2(z_{x}J_{x}+z_{y}J_{y})(1-2\varphi_{ij})-\ln{\left(1/\varphi_{ij}-1\right)},
    \label{main_mf_discrete}
\end{equation}
for $(i,j)\in(\{1,...,N^{x}_{b}\},\{1,...,N^{y}_{b}\})$.
Now we can carry out the scaling limit of the spin blocks, for which we
introduce the following notation: 
\begin{eqnarray}
    &{\rm lim}^{\mathbf{N}_{b}}_{\rm s}&\left[\varphi_{ij}=\varphi(il_{x},jl_{y})\right]\equiv\varphi(x,y), \ \forall (x,y) \in A, \nonumber \\ 
    &{\rm lim}^{\mathbf{N}_{b}}_{\rm s}&\left[\varphi_{i\pm1j}=\varphi(il_{x}\pm l_{x},j l_{y})\right]\equiv\lim\limits_{l_{x}\rightarrow0}\varphi(x\pm l_{x},y), \ \forall (x,y) \in A \nonumber \\
    &{\rm lim}^{\mathbf{N}_{b}}_{\rm s}&\left[\varphi_{ij\pm1}=\varphi(il_{x},j l_{y}\pm l_{y})\right]\equiv\lim\limits_{l_{y}\rightarrow0}\varphi(x,y\pm l_{y}), \ \forall (x,y) \in A,
    \label{varphi_x_def}
\end{eqnarray}
where $A=[-L_{x}/2,L_{x}/2]\times[-L_{y}/2,L_{y}/2]$. Applying ${\rm
  lim}^{\mathbf{N}_{b}}_{\rm s}\left[\cdot\right]$ to both sides of
Eq.~\eqref{main_mf_discrete} we obtain the following partial
differential equation
\begin{equation}
    \boxed{z_{x}J_{x}l^{2}_{x}\partial^{2}_{x}\varphi(x,y)+z_{y}J_{y}l^{2}_{y}\partial^{2}_{y}\varphi(x,y)=2(z_{x}J_{x}+z_{y}J_{y})(1-2\varphi(x,y))-\ln{\left(1/\varphi(x,y)-1\right)}, \ \forall (x,y) \in A,}
    \label{main_mf_continuous}
\end{equation}
where we used 
\begin{align*}
\lim_{l_{x}\rightarrow0}[\varphi(x+l_{x},y)-2\varphi(x,y)+\varphi(x-l_{x},y)]&=l^{2}_{x}\partial^{2}\varphi(x,y)/\partial x^{2}, \\
\lim_{l_{y}\rightarrow0}[\varphi(x,y+l_{y})-2\varphi(x,y)+\varphi(x,y-l_{y})]&=l^{2}_{y}\partial^{2}\varphi(x,y)/\partial y^{2}.
\end{align*}
Upon specifying the boundary conditions the solution to Eq.~\eqref{main_mf_continuous} gives the equilibrium concentration profile and maximizes the MF partition function in the thermodynamic limit. 
\subsection{Evaluating the partition function: Strategy 2}
\noindent To apply the thermodynamic limit of the spin blocks to Eq.~\eqref{fj_mf} we first add and subtract $z_{x}J_{x}\hat{\zeta}_{\rm MF}$ inside the third term. Next we use Eq.~\eqref{varphi_x_def} and obtain the following intermediate results
\begin{eqnarray}
    &\lim\limits_{l_{x}\rightarrow0}&[\hat{\zeta}_{\rm MF}(\varphi(x{+}l_{x},y),\varphi(x,y)){-}2\hat{\zeta}_{\rm MF}(\varphi(x,y),\varphi(x,y)){+}\hat{\zeta}_{\rm MF}(\varphi(x{-}l_{x},y),\varphi(x,y))]{=}l^{2}_{x}(1{-}\varphi(x,y))\partial^{2}_{x}\varphi(x,y), \nonumber \\
    &\lim\limits_{l_{y}\rightarrow0}&[\hat{\zeta}_{\rm MF}(\varphi(x,y{+}l_{y}),\varphi(x,y)){-}2\hat{\zeta}_{\rm MF}(\varphi(x,y),\varphi(x,y)){+}\hat{\zeta}_{\rm MF}(\varphi(x,y{-}l_{y}),\varphi(x,y))]{=}l^{2}_{y}(1{-}\varphi(x,y))\partial^{2}_{y}\varphi(x,y).
    \label{zeta_diff}
\end{eqnarray}
Inserting the outcome of Eq.~\eqref{zeta_diff} into Eq.~\eqref{fj_mf} we obtain the following result in the thermodynamic limit
\begin{equation}
    {\rm lim}^{\mathbf{N}_{b}}_{\rm s}\left[{\rm f}_{\rm MF}(\varphi_{i\pm1j},\varphi_{ij},\varphi_{ij\pm1})\right]={\rm f}_{\rm MF}(\varphi(x,y))+(1-\varphi(x,y))(z_{x}J_{x}l^{2}_{x}\partial^{2}_{x}\varphi(x,y)+z_{y}J_{y}l^{2}_{y}\partial^{2}_{y}\varphi(x,y))/2,
    \label{f_MF_per_site}
\end{equation}
where the MF local free energy density is given by
\begin{equation}
    {\rm f}_{\rm MF}(\varphi)\equiv\Xi(\varphi)+\Xi(1-\varphi)+2(z_{y}J_{y}+z_{x}J_{x})\left[\varphi(1-\varphi)-1/4\right].
    \label{f_uniform_MF_per_site}
\end{equation}
Finally, we construct the MF free energy functional which is given by (recall that $N_{b}=N^{x}_{b}\times N^{y}_{b}$)
\begin{eqnarray}
    F_{\rm MF}\left[\varphi(x,y)\right]&\equiv&{\rm lim}^{\mathbf{N}_{b}}_{\rm s}\left[(N_{b})^{-1}\sum_{i=1}^{N^{x}_{b}}\sum_{j=1}^{N^{y}_{b}}{\rm f}_{\rm MF}(\varphi_{i\pm1j},\varphi_{ij},\varphi_{ij\pm1})\right]\nonumber\\
    &=&\frac{1}{l_{x}l_{y}}\int\limits_{(x,y)\in A}[{\rm f}_{\rm MF}(\varphi(x,y))+(1-\varphi(x,y))(z_{x}J_{x}l^{2}_{x}\partial^{2}_{x}\varphi(x,y)+z_{y}J_{y}l^{2}_{y}\partial^{2}_{y}\varphi(x,y))/2]dxdy\nonumber\\
    &\stackrel{\rm P.I.}{=}&\frac{1}{l_{x}l_{y}}\int\limits_{(x,y)\in A}[{\rm f}_{\rm MF}(\varphi(x,y)){+}z_{x}J_{x}l^{2}_{x}(\partial_{x}\varphi(x,y))^{2}/2+z_{y}J_{y}l^{2}_{y}(\partial_{y}\varphi(x,y))^{2})/2]dxdy,
    \label{f_functional_MF}
\end{eqnarray}
where in the last line we carried out a partial integration (P.I.) and used zero-flux boundary conditions \\ $\partial_{y}\varphi(x,y)|_{y=\pm L_{y}/2}{=}\partial_{x}\varphi(x,y)|_{x=\pm L_{x}/2}=0$ which we will assume in later sections.
The profile $\varphi(x,y)$ which constitutes a stationary point of Eq.~\eqref{f_functional_MF}, i.e. $\delta F_{\rm MF}\left[\varphi(x,y)\right]/\delta \varphi(x,y)=0$, is obtained by solving the corresponding Euler-Lagrange equation
\begin{equation}
    \boxed{z_{x}J_{x}l^{2}_{x}\partial^{2}_{x}\varphi(x,y)+z_{y}J_{y}l^{2}_{y}\partial^{2}_{y}\varphi(x,y)=\partial_{\varphi(x,y)}{\rm f}_{\rm MF}(\varphi(x,y)), \ \forall (x,y) \in A.}
    \label{Euler-Lagrange-MF}
\end{equation}
Plugging Eq.~\eqref{f_uniform_MF_per_site} into Eq.~\eqref{Euler-Lagrange-MF} finally results in Eq.~\eqref{main_mf_continuous}. 
\section{Bethe-Guggenheim approximation}\label{secV}
\subsection{Introduction}
Our starting point within the BG approximation is the free energy density given by Eq.~(26) in the main article, which reads (for a detailed derivation see Sec.~V in the main article)
\begin{align}
    &{\rm f}(\varphi_{i\pm1j},\varphi_{ij},\varphi_{ij\pm1})\equiv{\rm lim}^{\mathbf{N}_{\sigma}}_{\rm s}\left[-(N^{b}_{\sigma})^{-1}\ln{(\Psi(\varphi_{ij},\hat{\zeta}^{x,y}_{\dagger},\hat{\xi}^{x,y\pm}_{\dagger})\e{-N^{b}_{\sigma}[z_{x}J_{x}(\hat{\zeta}^{x}_{\dagger}+\{\hat{\xi}^{x+}_{\dagger}+\hat{\xi}^{x-}_{\dagger}\}/2)+(x\leftrightarrow y)-\mathcal{C}]})}\right] \nonumber \\
    &=(z_{x}/8)\textstyle{\sum_{\pm}}[\Xi(1-\varphi_{ij}-\hat{\xi}^{x\pm}_{\dagger})+\Xi(\varphi_{i\pm1j}-\hat{\xi}^{x\pm}_{\dagger})+\Xi(\varphi_{ij}-\varphi_{i\pm1j}+\hat{\xi}^{x\pm}_{\dagger})+\Xi(\hat{\xi}^{x\pm}_{\dagger})-\Xi(1-\varphi_{i\pm1j})-\Xi(\varphi_{i\pm1j})]\nonumber\\
    &+(z_{y}/8)\textstyle{\sum_{\pm}}[\Xi(1-\varphi_{ij}-\hat{\xi}^{y\pm}_{\dagger})+\Xi(\varphi_{ij\pm1}-\hat{\xi}^{y\pm}_{\dagger})+\Xi(\varphi_{ij}-\varphi_{ij\pm1}+\hat{\xi}^{y\pm}_{\dagger})+\Xi(\hat{\xi}^{y\pm}_{\dagger})-\Xi(1-\varphi_{ij\pm1})-\Xi(\varphi_{ij\pm1})]\nonumber\\
    &+\{(z_{x}/4)[\Xi(1-\varphi_{ij}-\hat{\zeta}^{x}_{\dagger})+\Xi(\varphi_{ij}-\hat{\zeta}^{x}_{\dagger})+2\Xi(\hat{\zeta}^{x}_{\dagger})]+(x\leftrightarrow y)\}+(1-3z/4)[\Xi(\varphi_{ij})+\Xi(1-\varphi_{ij})]\nonumber\\
    &+\{z_{x}J_{x}(\hat{\zeta}^{x}_{\dagger}+\{\hat{\xi}^{x+}_{\dagger}+\hat{\xi}^{x-}_{\dagger}\}/2)+(x\leftrightarrow y)\}-\mathcal{C},
    \label{fj_BG2}
\end{align}
where $\Xi(x)\equiv x\ln{(x)}$ and the functions $(\hat{\zeta}^{x,y}_{\dagger},\hat{\xi}^{x,y\pm}_{\dagger})$ are given by Eq.~(25) in the main article. Comparing Eq.~\eqref{fj_BG2} with
Eq.~\eqref{fj_mf} we notice that the BG free energy density has
considerably more terms than its MF counterpart due to the functional
form of the degeneracy factor. As with the MF calculation we will
consider two different strategies for carrying out the optimization over
$\varphi_{ij}$. The second strategy has already been discussed in the main article, and here we provide some further details about the calculation. 
\subsection{Evaluating the partition function: Strategy 1}
\noindent The local minima of Eq.~\eqref{fj_BG2} w.r.t\ $\varphi_{ij}$ are given by the following equation
\begin{equation}
    \partial_{\varphi_{ij}}[{\rm f}(\varphi_{i\pm1j},\varphi_{ij},\varphi_{ij\pm1})+\sum_{k=\pm1}\left({\rm f}(\varphi_{i+k\pm1j},\varphi_{i+kj},\varphi_{i+kj\pm1})+{\rm f}(\varphi_{i\pm1j+k},\varphi_{ij+k},\varphi_{ij+k\pm1})\right)]\stackrel{!}{=}0.
    \label{Optim_BG}
\end{equation}
Upon taking the partial derivative of the BG local free energy density w.r.t.\ $\varphi_{ij}$, we can use the following
\begin{equation}
    \partial_{\hat{\zeta}^{x,y}_{\dagger}}{\rm f}(\varphi_{i\pm1j},\varphi_{ij},\varphi_{ij\pm1})= \partial_{\hat{\xi}^{x,y\pm}_{\dagger}}{\rm f}(\varphi_{i\pm1j},\varphi_{ij},\varphi_{ij\pm1})=0,
\end{equation}
since both $\hat{\zeta}^{x,y}_{\dagger}$ and $\hat{\xi}^{x,y\pm}_{\dagger}$ are obtained by minimization of the BG free energy density. This renders the evaluation of Eq.~\eqref{Optim_BG} a relatively easy task and results in the following recurrent set of difference equations
\begin{eqnarray}
    &\textcolor{white}{.}&\frac{z_{x}}{8}\sum_{\pm}\left[\ln{\left(\frac{1{-}\varphi_{ij}{-}\hat{\xi}^{x\pm}_{\dagger}(\varphi_{i\pm1j},\varphi_{ij})}{\varphi_{ij}{-}\varphi_{i\pm1j}{+}\hat{\xi}^{x\pm}_{\dagger}(\varphi_{i\pm1j},\varphi_{ij})}\right)}-\ln{
    \left(\frac{\varphi_{ij}{-}\hat{\xi}^{x\pm}_{\dagger}(\varphi_{ij},\varphi_{i\pm1j})}{\varphi_{i\pm1j}{-}\varphi_{ij}{+}\hat{\xi}^{x\pm}_{\dagger}(\varphi_{ij},\varphi_{i\pm1j})}\right)
    }\right]+\nonumber\\
    &\textcolor{white}{.}&\frac{z_{y}}{8}\sum_{\pm}\left[\ln{\left(\frac{1{-}\varphi_{ij}{-}\hat{\xi}^{y\pm}_{\dagger}(\varphi_{ij\pm1},\varphi_{ij})}{\varphi_{ij}{-}\varphi_{ij\pm1}{+}\hat{\xi}^{y\pm}_{\dagger}(\varphi_{ij\pm1},\varphi_{ij})}\right)}-\ln{
    \left(\frac{\varphi_{ij}{-}\hat{\xi}^{y\pm}_{\dagger}(\varphi_{ij},\varphi_{ij\pm1})}{\varphi_{ij\pm1}{-}\varphi_{ij}{+}\hat{\xi}^{y\pm}_{\dagger}(\varphi_{ij},\varphi_{ij\pm1})}\right)
    }\right]\nonumber\\
    &\textcolor{white}{.}&=\frac{z_{x}}{4}\ln{\left(\frac{\varphi_{ij}{-}\hat{\zeta}^{x}_{\dagger}(\varphi_{ij},\varphi_{ij})}{1{-}\varphi_{ij}{-}\hat{\zeta}^{x}_{\dagger}(\varphi_{ij},\varphi_{ij})}\right)}{+}(x\leftrightarrow y){+}(1{-}z)\ln{\left(\frac{\varphi_{ij}}{1{-}\varphi_{ij}}\right)}{-}\mu, \ \forall (i,j) \in (\{1,...,N^{x}_{b}\},\{1,...,N^{y}_{b}\}).
    \label{main_bg_discrete}
\end{eqnarray}
For a one-dimensional concentration profile (i.e.\ $\varphi_{ij}\rightarrow\varphi_{i}$) a similar
equation has been derived in \cite{sParlange1968} -- see Eqs.~(31)-(33)
therein -- where the solution is obtained (only) around the critical point. Here we proceed with applying the thermodynamic limit of the spin blocks to Eq.~\eqref{main_bg_discrete} using Eq.~\eqref{varphi_x_def}. To obtain the thermodynamic limit we calculate the following terms:
\begin{eqnarray}
    \lim\limits_{l_{x}\rightarrow0}[(\textstyle{\sum_{\pm}}\ln(1{-}\varphi(x,y){-}\hat{\xi}^{x\pm}_{\dagger}(\varphi(x{\pm}l_{x},y),\varphi(x,y)))\textcolor{blue}{{-}2\ln(1{-}\varphi(x,y){-}\hat{\zeta}^{x}_{\dagger}(\varphi(x,y),\varphi(x,y))})/l^{2}_{x}]=\nonumber\\
    \left(\frac{\hat{\xi}^{x(1,0)}_{\dagger}}{\varphi(x,y)+\hat{\zeta}^{x}_{\dagger}-1}\right)\partial^{2}_{x}\varphi(x,y)-\left(\frac{(\hat{\xi}^{x(1,0)}_{\dagger})^2}{(\varphi(x,y)+\hat{\zeta}^{x}_{\dagger}-1)^2}-\frac{\hat{\xi}^{x(2,0)}_{\dagger}}{\varphi(x,y)+\hat{\zeta}^{x}_{\dagger}-1}\right)(\partial_{x}\varphi(x,y))^{2},
    \label{rela1}
\end{eqnarray}
\begin{eqnarray}
    \lim\limits_{l_{x}\rightarrow0}[(\textcolor{blue}{2\ln(\varphi(x,y)-\hat{\zeta}^{x}_{\dagger}(\varphi(x,y),\varphi(x,y)))}-\textstyle{\sum_{\pm}}\ln(\varphi(x,y)-\hat{\xi}^{x\pm}_{\dagger}(\varphi(x,y),\varphi(x\pm l_{x},y))))/l^{2}_{x}]=\nonumber\\
    \left(\frac{\hat{\xi}^{x(0,1)}_{\dagger}}{\varphi(x,y)-\hat{\zeta}^{x}_{\dagger}}\right)\partial^{2}_{x}\varphi(x,y)+\left(\frac{(\hat{\xi}^{x(0,1)}_{\dagger})^2}{(\varphi(x,y)-\hat{\zeta}^{x}_{\dagger})^2}+\frac{\hat{\xi}^{x(0,2)}_{\dagger}}{\varphi(x,y)-\hat{\zeta}^{x}_{\dagger}}\right)(\partial_{x}\varphi(x,y))^{2},
    \label{rela2}
\end{eqnarray}
\begin{eqnarray}
    \lim\limits_{l_{x}\rightarrow0}[(\textstyle{\sum_{\pm}}\ln(\varphi(x\pm l_{x},y)-\varphi(x,y)+\hat{\xi}^{x\pm}_{\dagger}(\varphi(x,y),\varphi(x\pm l_{x},y)))\textcolor{purple}{-2\ln(\hat{\zeta}^{x}_{\dagger}(\varphi(x,y),\varphi(x,y)))})/l^{2}_{x}]=\nonumber\\
    \left(\frac{\hat{\xi}^{x(0,1)}_{\dagger}+1}{\hat{\zeta}^{x}_{\dagger}}\right)\partial^{2}_{x}\varphi(x,y)-\left(\frac{(\hat{\xi}^{x(0,1)}_{\dagger}+1)^2}{(\hat{\zeta}^{x}_{\dagger})^2}-\frac{\hat{\xi}^{x(0,2)}_{\dagger}}{\hat{\zeta}^{x}_{\dagger}}\right)(\partial_{x}\varphi(x,y))^{2},
    \label{rela3}
\end{eqnarray}
\begin{eqnarray}
    \lim\limits_{l_{x}\rightarrow0}[(\textcolor{purple}{2\ln(\hat{\zeta}^{x}_{\dagger}(\varphi(x,y),\varphi(x,y)))}-\textstyle{\sum_{\pm}}\ln(\varphi(x,y)-\varphi(x\pm l_{x},y)+\hat{\xi}^{x\pm}_{\dagger}(\varphi(x\pm l_{x},y),\varphi(x,y))))/l^{2}_{x}]=\nonumber\\
    \left(\frac{1-\hat{\xi}^{x(1,0)}_{\dagger}}{\hat{\zeta}^{x}_{\dagger}}\right)\partial^{2}_{x}\varphi(x,y)+\left(\frac{(\hat{\xi}^{x(1,0)}_{\dagger}-1)^2}{(\hat{\zeta}^{x}_{\dagger})^2}-\frac{\hat{\xi}^{x(2,0)}_{\dagger}}{\hat{\zeta}^{x}_{\dagger}}\right)(\partial_{x}\varphi(x,y))^{2},
    \label{rela4}
\end{eqnarray}
where $\hat{\xi}^{x(m,n)}_{\dagger}\equiv\partial^{m}_{a}\partial^{n}_{b}\hat{\xi}^{x}_{\dagger}(a,b)|_{(\varphi(x,y),\varphi(x,y))}$ and we have used that $\hat{\xi}^{x\pm}_{\dagger}(a,a)=\hat{\zeta}^{x}_{\dagger}(a,a)$. Upon interchanging $x$ with $y$ the results of Eqs.~\eqref{rela1}-\eqref{rela4} also apply to the $y$-direction. Note that the \textcolor{blue}{blue} terms in Eqs.~\eqref{rela1} and \eqref{rela2} are added manually, and therefore also need to be added to the RHS of Eq.~\eqref{main_bg_discrete}. The \textcolor{purple}{purple} terms in Eqs.~\eqref{rela3} and \eqref{rela4} directly cancel, and therefore do not need to be added to the RHS. Summing up all the contributions we obtain the following expression
\begin{equation}
    \frac{z_{x}}{8}(\eqref{rela1}+\eqref{rela2}+\eqref{rela3}+\eqref{rela4})=\kappa_{x}(\varphi(x,y))\partial^{2}_{x}\varphi(x,y)+\frac{\kappa^{\prime}_{x}(\varphi(x,y))(\partial_{x}\varphi(x,y))^{2}}{2},
\end{equation}
where the gradient energy coefficient $\kappa_{x}$ is given by 
\begin{equation}
    \kappa_{(x,y)}(\varphi)\equiv\frac{z_{x,y}(\e{4J_{x,y}}-1)}{4\sqrt{1+4(\e{4J_{x,y}}-1)\varphi(1-\varphi)}},
    \label{kappaSM}
\end{equation}
and $\kappa_{
  x}^{\prime}(\varphi)=\partial_{\varphi}\kappa_{{\rm BG},
  x}(\varphi)$. Eq.~\eqref{kappaSM} is also given in the main article as
Eq.~(29). For a one-dimensional concentration profile (only)
this result has also been derived in
\cite{skikuchi1962theory2} -- see Eq.~(2.12b) therein -- but so far it has not
been derived for a two-dimensional system. 
Plugging the result back into the LHS of Eq.~\eqref{main_bg_discrete}
we obtain the following partial differential equation (PDE)
\begin{equation}
    \boxed{l^{2}_{x}[\kappa_{x}\partial^{2}_{x}\varphi(x,y){+}\kappa^{\prime}_{x}(\partial_{x}\varphi(x,y))^{2}/2]{+}(x\leftrightarrow y){=}\frac{z_{x}}{2}\ln{\left(\frac{\varphi(x,y){-}\hat{\zeta}^{x}_{\dagger}}{1{-}\varphi(x,y){-}\hat{\zeta}^{x}_{\dagger}}\right)}{+}(x\leftrightarrow y){+}(1{-}z)\ln{\left(\frac{\varphi(x,y)}{1{-}\varphi(x,y)}\right)}{-}\mu,}
    \label{main_bg_continuous}
\end{equation}
which applies in the domain $(x,y)\in A$ with $A=[-L_{x}/2,L_{x}/2]\times[-L_{y}/2,L_{y}/2]$. Recall that $(x\leftrightarrow y)$ denotes a repetition of the preceding term with $x$ substituted by $y$, and $\hat{\zeta}^{x,y}_{\dagger}$ is given by the first equation in Eq.~(25) in the main manuscript. Eq.~\eqref{main_bg_continuous} is the BG equivalent of the MF PDE given by Eq.~\eqref{main_mf_continuous}.
\subsection{Evaluating the partition function: Strategy 2}
\noindent Applying the thermodynamic limit to Eq.~\eqref{fj_BG2} in the $x$-direction we need to keep track of the following terms:
\begin{equation}
    \lim\limits_{l_{x}\rightarrow0}[(\textstyle{\sum_{\pm}}\hat{\xi}^{x\pm}_{\dagger}(\varphi(x{\pm} l_{x},y),\varphi(x,y)){-}2\hat{\zeta}^{x}_{\dagger}(\varphi(x,y),\varphi(x,y)))/l^{2}_{x}]
    \stackrel{\rm P.I.}{=}
    -(\partial_{x}\varphi(x,y))^{2}\hat{\xi}^{x(1,1)}_{\dagger},
    \label{rel1}
\end{equation}
\begin{equation}
    \lim\limits_{l_{x}\rightarrow0}[(\textstyle{\sum_{\pm}}\Xi(\varphi(x\pm l_{x},y))-2\Xi(\varphi(x,y)))/l^{2}_{x}]=(\partial_{x}\varphi(x,y))^{2}/\varphi(x,y)+\partial^{2}_{x}\varphi(x,y)(\ln{(\varphi(x,y))}+1)\stackrel{\rm P.I.}{=}0,
    \label{rel2}
\end{equation}
\begin{equation}
    \lim\limits_{l_{x}\rightarrow0}[(\textstyle{\sum_{\pm}}\Xi(1{-}\varphi(x\pm l_{x},y)){-}2\Xi(1{-}\varphi(x,y)))/l^{2}_{x}]{=}(\partial_{x}\varphi(x,y))^{2}/(1{-}\varphi(x,y)){-}\partial^{2}_{x}\varphi(x,y)(\ln{(1{-}\varphi(x,y))}{+}1)\stackrel{\rm P.I.}{=}0,
    \label{rel3}
\end{equation}
\begin{eqnarray}
    \lim\limits_{l_{x}\rightarrow0}[(\textstyle{\sum_{\pm}}\Xi(\hat{\xi}^{x\pm}_{\dagger}(\varphi(x\pm l_{x},y),\varphi(x,y)))-2\Xi(\hat{\zeta}^{x}_{\dagger}(\varphi(x,y),\varphi(x,y))))/l^{2}_{x}]
    \stackrel{\rm P.I.}{=}\nonumber\\
    \textcolor{purple}{-(\partial_{x}\varphi(x,y))^{2}\frac{\hat{\xi}^{x(1,0)}_{\dagger}\hat{\xi}^{x(0,1)}_{\dagger}}{\hat{\zeta}^{x}_{\dagger}}}
    \textcolor{blue}{-(\partial_{x}\varphi(x,y))^{2}\hat{\xi}^{x(1,1)}_{\dagger}(\ln{(\hat{\zeta}^{x}_{\dagger})}+1)},
    \label{rel4}
\end{eqnarray}
\begin{eqnarray}
    \lim\limits_{l_{x}\rightarrow0}[(\textstyle{\sum_{\pm}}\Xi(\varphi(x\pm l_{x},y){-}\hat{\xi}^{x\pm}_{\dagger}(\varphi(x\pm l_{x},y),\varphi(x,y)))-2\Xi(\varphi(x,y)-\hat{\zeta}^{x}_{\dagger}(\varphi(x,y),\varphi(x,y))))/l^{2}_{x}]\stackrel{\rm P.I.}{=}\nonumber\\
    \textcolor{purple}{-(\partial_{x}\varphi(x,y))^{2}\frac{(\hat{\xi}^{x(1,0)}_{\dagger}-1)\hat{\xi}^{x(0,1)}_{\dagger}}{\varphi(x,y)-\hat{\zeta}^{x}_{\dagger}}}
    \textcolor{blue}{+(\partial_{x}\varphi(x,y))^{2}\hat{\xi}^{x(1,1)}_{\dagger}(\ln{(\varphi(x,y)-\hat{\zeta}^{x}_{\dagger})}+1)},
    \label{rel5}
\end{eqnarray}
\begin{eqnarray}
    \lim\limits_{l_{x}\rightarrow0}[(\textstyle{\sum_{\pm}}\Xi(1-\varphi(x,y)-\hat{\xi}^{x\pm}_{\dagger}(\varphi(x\pm l_{x},y),\varphi(x,y)))-2\Xi(1-\varphi(x,y)-\hat{\zeta}^{x}_{\dagger}(\varphi(x,y),\varphi(x,y))))/l^{2}_{x}]\stackrel{\rm P.I.}{=}\nonumber\\
    \textcolor{purple}{-(\partial_{x}\varphi(x,y))^{2}\frac{\hat{\xi}^{x(1,0)}_{\dagger}[\hat{\xi}^{x(0,1)}_{\dagger}{+}1]}{1-\varphi(x,y)-\hat{\zeta}^{x}_{\dagger}}}
    \textcolor{blue}{+(\partial_{x}\varphi(x,y))^{2}\hat{\xi}^{x(1,1)}_{\dagger}(\ln{(1-\varphi(x,y)-\hat{\zeta}^{x}_{\dagger})}+1)},
    \label{rel6}
\end{eqnarray}
\begin{eqnarray}
    \lim\limits_{l_{x}\rightarrow0}[(\textstyle{\sum_{\pm}}\Xi(\varphi(x,y){-}\varphi(x\pm l_{x},y){+}\hat{\xi}^{x\pm}_{\dagger}(\varphi(x\pm l_{x},y),\varphi(x,y)))-2\Xi(\hat{\zeta}^{x}_{\dagger}(\varphi(x,y),\varphi(x,y))))/l^{2}_{x}]\stackrel{\rm P.I.}{=}\nonumber\\
    \textcolor{purple}{-(\partial_{x}\varphi(x,y))^{2}\frac{(\hat{\xi}^{x(1,0)}_{\dagger}-1)(\hat{\xi}^{x(0,1)}_{\dagger}+1)}{\hat{\zeta}^{x}_{\dagger}}}
    \textcolor{blue}{-(\partial_{x}\varphi(x,y))^{2}\hat{\xi}^{x(1,1)}_{\dagger}(\ln{(\hat{\zeta}^{x}_{\dagger})}+1)},
    \label{rel7}
\end{eqnarray}
where we have immediately carried out a partial integration -- since each term will arise inside an integral -- 
and used zero-flux boundary conditions $\partial_{x}\varphi(x,y)|_{x=\pm L_{x}/2}=0$ to express everything in terms of $(\partial_{x}\varphi(x,y))^{2}$. 
Next we add up all the \textcolor{blue}{blue} terms in Eqs.~\eqref{rel4}-\eqref{rel7} and find that they exactly cancel with Eq.~\eqref{rel1} upon plugging them back into Eq.~\eqref{fj_BG2}. Adding up all the \textcolor{purple}{purple} terms in Eqs.~\eqref{rel4}-\eqref{rel7} gives the following result
\begin{equation}
    \frac{z_{x}}{8}(\textcolor{purple}{\eqref{rel4}+\eqref{rel5}+\eqref{rel6}+\eqref{rel7}})=\frac{1}{2}\kappa_{x}(\varphi(x,y))(\partial_{x}\varphi(x,y))^{2}.
\end{equation}
Upon interchanging $x$ with $y$ the same results applies to the
$y$-direction. Putting the results back into Eq.~\eqref{fj_BG2} and
adding/subtracting those terms which have been added by hand in
Eqs.~\eqref{rel1}-\eqref{rel7} we finally obtain the BG free energy
density in the scaling limit of the blocks
\begin{equation}
    {\rm lim}^{\mathbf{N}_{b}}_{\rm s}\left[{\rm f}(\varphi_{i\pm1j},\varphi_{ij},\varphi_{ij\pm1})\right]=
    {\rm f}(\varphi(x,y))+\frac{l^{2}_{x}}{2}\kappa_{x}(\varphi(x,y))(\partial_{x}\varphi(x,y))^{2}+\frac{l^{2}_{y}}{2}\kappa_{y}(\varphi(x,y))(\partial_{y}\varphi(x,y))^{2},
    \label{f_BG_per_site}
\end{equation}
where $\kappa_{x,y}(\varphi)$ is defined in Eq.~\eqref{kappaSM}, and the BG local free energy density ${\rm f}(\varphi)$ is given by Eq.~(28) in the main article. Finally, the BG free energy density functional is given by
\begin{eqnarray}
    F\left[\varphi(x,y)\right]&\equiv&{\rm lim}^{\mathbf{N}_{b}}_{\rm s}\left[(N_{b})^{-1}\sum_{i=1}^{N^{x}_{b}}\sum_{j=1}^{N^{y}_{b}}{\rm f}(\varphi_{i\pm1j},\varphi_{ij},\varphi_{ij\pm1})\right]\nonumber\\
    &=&\frac{1}{l_{x}l_{y}}\int\limits_{(x,y)  \in A}[{\rm f}(\varphi(x,y))+\frac{l^{2}_{x}}{2}\kappa_{x}(\partial_{x}\varphi(x,y))^{2}+\frac{l^{2}_{y}}{2}\kappa_{y}(\partial_{y}\varphi(x,y))^{2}]dxdy,
    \label{f_functional_BG}
\end{eqnarray}
which is also reported in the main article as Eq.~(27).
Note that the coordinates $\mathbf{x}$ in the main article have been written  in units of the box size $(l_{x},l_{y})$, which is equivalent to setting $l_{x}=l_{y}=1$ in Eq.~\eqref{f_functional_BG}.
The profile $\varphi(x,y)$ which constitutes a stationary point of Eq.~\eqref{f_functional_BG}, i.e. $\delta F\left[\varphi(x,y)\right]/\delta \varphi(x,y)=0$, is obtained by solving the corresponding Euler-Lagrange equation 
\begin{equation}
    \boxed{l^{2}_{x}[\kappa_{x}\partial^{2}_{x}\varphi(x,y)+\kappa^{\prime}_{x}(\partial_{x}\varphi(x,y))^{2}/2]+(x\leftrightarrow y)=\partial_{\varphi(x,y)} {\rm f}(\varphi(x,y)).}
    \label{Euler-Lagrange-BG}
\end{equation}
which is equivalent to Eq.~\eqref{main_bg_continuous}.
\section{Equilibrium concentration profile}\label{secVI}
\noindent Here we consider a
concentration profile which only varies
in the $x$ direction,
i.e.\ $\varphi(x,y)=\varphi(x), \ \forall
\ x\in[-L_{x}/2,L_{x}/2]$. The equilibrium profile
$\varphi(x)$ is an extremum of Eqs.~\eqref{f_functional_MF} and
\eqref{f_functional_BG} for the MF and BG approximation,
respectively. Here we will derive analytical expressions for the
interfacial steepness, interfacial width (according to the Cahn-Hilliard definition), and prove the broadening of the BG equilibrium profile.
\subsection{Results within Mean Field theory}
\begin{figure}
    \centering
    \includegraphics[width=0.95\textwidth]{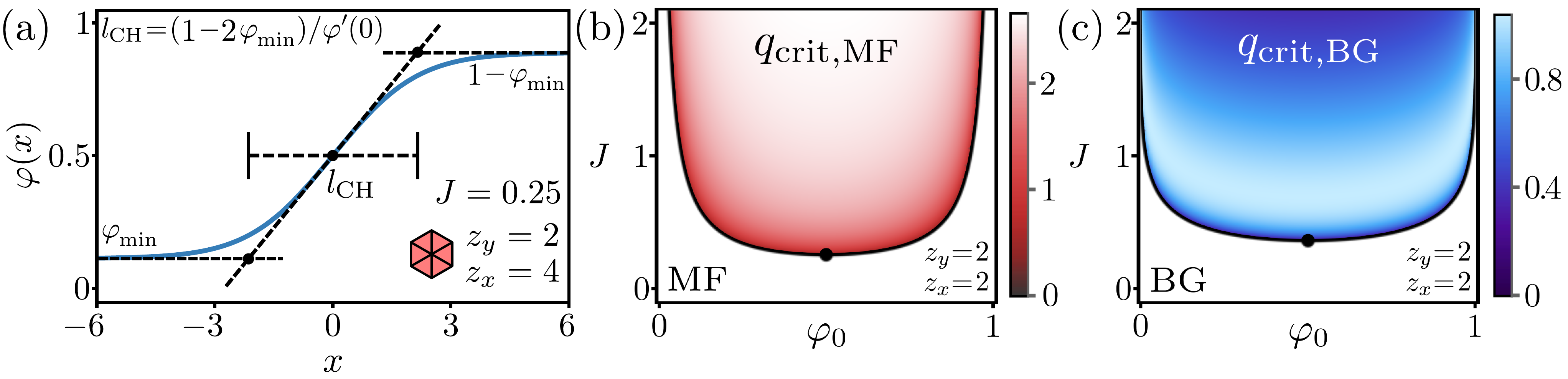}
    \caption{(a) Representation of the Cahn-Hilliard interfacial width $l_{\rm CH}$ used in Eq.~\eqref{lch-MF}. Here we used the concentration profile for a hexagonal ($z=6$) lattice obtained with the BG approximation. (b) Critical wavevector obtained with the MF approximation Eq.~\eqref{qcrit-mf} for a square lattice. The black line represents the MF spinodal and the black dot the MF critical point $J_{\rm crit, MF}=1/4$. (c) Critical wavevector obtained with the BG approximation Eq.~\eqref{qc-BG2} for a square lattice. The black line represents the BG spinodal and the black dot the BG critical point $J_{\rm crit, BG}=\ln{(2)}/2$. }
    \label{fig1SM}
\end{figure}

\noindent For a one-dimensional concentration profile Eq.~\eqref{Euler-Lagrange-MF} reduces to a second order autonomous ODE. Therefore we can directly obtain the interfacial steepness $\varphi_{\rm MF}^{\prime}(x)$, which reads
\begin{equation}
    \varphi_{\rm MF}^{\prime}(x)=\pm\sqrt{2({\rm f}_{\rm MF}(\varphi_{\rm MF}(x))-{\rm f}_{\rm MF}(\varphi_{\rm MF, min}))/z_{x}J_{x}},
    \label{dphi-MF}
\end{equation}
where we have set the integration constant to
$\mathcal{C}_{1}=-{\rm f}_{\rm MF}(\varphi_{\rm MF, min})$
with $\varphi_{\rm MF, min}\equiv\arg {\rm inf}_{0\leq\varphi\leq1/2}
\ {\rm f}_{\rm MF}(\varphi)$ such that the term inside the
square root on the RHS is always positive and
$\lim_{x\rightarrow\pm\infty}\varphi_{\rm MF}^{\prime}(x)=0$. The
location of the global minimum of the uniform MF free energy density
can be written as $\varphi_{\rm MF, min}\equiv (1- |s|)/2$, where $s\in[-1,1]$ is given by the nonzero solutions to the so-called transcendental mean field equation \cite{sblom2022criticality}
\begin{equation}
    s=\tanh{([z_{x}J_{x}+z_{y}J_{y}]s)}.
    \label{MF-equation}
\end{equation}
Below the critical coupling for $z_{x}J_{x}+z_{y}J_{y}\leq1$ and the only solution to Eq.~\eqref{MF-equation} is given by $s=0$, resulting in $\varphi_{\rm MF, min}=1/2$. Above the critical coupling for $z_{x}J_{x}+z_{y}J_{y}>1$ there exists two nonzero solutions resulting in $\varphi_{\rm MF, min}<1/2$. Now let us focus on the isotropic case with a vanishing external field, i.e.\ $J_{x}=J_{y}=J$, and consider the interfacial steepness at $x=0$. Based on the imposed boundary conditions we know that $\varphi_{\rm MF}(0)=1/2$, and therefore the interfacial steepness at $x=0$ reads
\begin{equation}
    \varphi_{\rm MF}^{\prime}(0)=\pm\sqrt{2({\rm f}_{\rm MF}(1/2)-{\rm f}_{\rm MF}(\varphi_{\rm MF, min}))/z_{x}J}.
    \label{dphi-MF-0}
\end{equation}
For a square lattice ($\{z_{x},z_{y}\}=\{2,2\}$) Eq.~\eqref{dphi-MF-0} is shown in Fig.~4b in the main article with the red solid line. 
\\
\\
\noindent To obtain the interfacial width as defined by Cahn and Hilliard (see Eq.~(2.25) in \cite{scahn1958freeI2}) we simply need take a line tangential to the slope of the concentration profile at $x=0$ and determine the crossing points of this line with the bulk concentration values as depicted in Fig.~\ref{fig1SM}a. This leads to the expression
\begin{equation}
    l_{\rm MF, CH}=(1-2\varphi_{\rm MF, min})/\varphi_{\rm MF}^{\prime}(0),
    \label{lch-MF}
\end{equation}
where we insert Eq.~\eqref{dphi-MF-0} for $\varphi_{\rm MF}^{\prime}(0)$ with the positive sign. We see that for $zJ\leq1$ we have $l_{\rm MF, CH}\rightarrow\infty$. Now let us consider an infinite coupling strength. In this limit the nonzero solutions to Eq.~\eqref{MF-equation} are trivially given by $s=\pm 1, \ \forall z>0$, and therefore we obtain
\begin{equation}
    \lim\limits_{J\rightarrow\infty}\varphi_{\rm MF}^{\prime}(0)=\lim\limits_{J\rightarrow\infty}\pm\sqrt{2({\rm f}_{\rm MF}(1/2)-{\rm f}_{\rm MF}(0))/z_{x}J}{=}\lim\limits_{J\rightarrow\infty}\pm\sqrt{2(zJ/2-\ln{(2)})/z_{x}J}=\pm\sqrt{z/z_{x}}.
\end{equation}
Hence, in the infinite coupling limit the interfacial steepness converges to a maximum finite nonzero value. This value for the MF interfacial steepness is also reported in Fig.~4b in the main article. Furthermore, the interfacial width decreases and converges to the value
\begin{equation}
    \lim\limits_{J\rightarrow\infty}l_{\rm MF, CH}=\lim\limits_{J\rightarrow\infty}(1-2\varphi_{\rm MF, min})/\varphi_{\rm MF}^{\prime}(0)=\sqrt{z_{x}/z}.
\end{equation}
\subsection{Results within Bethe-Guggenheim theory}
\noindent Similar to the MF analysis Eq.~\eqref{Euler-Lagrange-BG} reduces to a
second order autonomous ODE for a one-dimensional concentration profile. To obtain the interfacial steepness we first rewrite the LHS of Eq.~\eqref{Euler-Lagrange-BG} as
\begin{equation}
    \kappa_{x}\varphi^{\prime\prime}(x)+\kappa^{\prime}_{x}(\varphi^{\prime}(x))^{2}/2=\frac{1}{2\varphi^{\prime}(x)}\frac{d}{dx}[\kappa_{x}(\varphi^{\prime}(x))^{2}].
\end{equation}
Taking the term $1/\varphi^{\prime}(x)$ to the RHS of Eq.~\eqref{Euler-Lagrange-BG} and using the fact that $\varphi^{\prime}(x)(\partial {\rm f}(\varphi(x))/\partial \varphi(x))=d {\rm f}(\varphi(x))/dx$, we can integrate both sides over $x$, resulting in the first-order autonomous ODE
\begin{equation}
    \frac{1}{2}\kappa_{x}(\varphi^{\prime}(x))^{2}={\rm f}(\varphi(x))+\mathcal{C}_{1},
    \label{Euler-Lagrange-BG2}
\end{equation}
where $\mathcal{C}_{1}$ is an integration constant. From Eq.~\eqref{Euler-Lagrange-BG2} we can directly readout the interfacial steepness
\begin{equation}
    \varphi^{\prime}(x)=\pm\sqrt{2({\rm f}(\varphi(x))-{\rm f}(\varphi_{\rm min}))/\kappa_{x}(\varphi(x))},
    \label{dphi-BG}
\end{equation}
where we have set the integration constant $\mathcal{C}_{1}=-{\rm f}(\varphi_{\rm min})$ with $\varphi_{\rm min}\equiv{\rm inf}_{0\leq\varphi\leq1/2} \ {\rm f}(\varphi)$. The integration constant is chosen such that the term inside the square root on the RHS is always positive and to impose a vanishing derivative at the boundaries. Now let us focus specifically on the isotropic case with a vanishing external field, i.e.\ $J_{x}=J_{y}=J$. The location of the global minimum $\varphi_{\rm min}$ of the local BG free energy density can be written as $\varphi_{\rm min}=\chi_{\varphi}/(1+\chi_{\varphi})$, where $\chi_{\varphi}\in[0,\infty)$ is given by the nontrivial solutions (i.e. $\chi_{\varphi}\neq1$) to the transcendental equation \cite{sblom2022criticality}
\begin{equation}
    \chi_{\varphi}-\e{2J}[\e{\mu/z}\chi^{(z-1)/z}_{\varphi}-\e{-\mu/z}\chi^{1/z}_{\varphi}]-1=0.
    \label{transcendental-eq-BG}
\end{equation}
Below and at the critical coupling $J\leq\ln{(z/(z-2))}/2$  Eq.~\eqref{transcendental-eq-BG} has one trivial solution $\chi_{\varphi}=1$, resulting in $\varphi_{\rm min}=1/2$. Above the critical coupling there exists two nontrivial solutions, resulting in $\varphi_{\rm min}<1/2$. Eq.~\eqref{transcendental-eq-BG} cannot be solved analytically for general $z$ but is explicitly solvable for a triangular, square, and hexagonal lattice, which gives
\begin{eqnarray}
    \varphi_{\rm min}|_{z=3}&=&
    \begin{dcases*}
        \frac{1}{2} & \hspace{3.85cm} , $0\leq J \leq  \ln{(3)}/2$\\ 
        \frac{1}{2}\left[1-\frac{\e{2J}((\e{2J}+1)(\e{2J}-3))^{\frac{1}{2}}}{2(\e{3J}\sinh{(J)}-1)}\right] & \hspace{3.85cm} , $J \geq   \ln{(3)}/2$
    \end{dcases*}, \nonumber \\
    \varphi_{\rm min}|_{z=4}&=&
    \begin{dcases*}
        \frac{1}{2} & \hspace{5.5cm} , $0\leq J \leq  \ln{(2)}/2$\\ 
        \frac{1}{2}\left[1-\frac{\e{2J}(\e{4 J}-4)^{\frac{1}{2}}}{\e{4J}-2}\right] & \hspace{5.5cm} , $J \geq   \ln{(2)}/2$
    \end{dcases*},\nonumber\\
    \varphi_{\rm min}|_{z=6}&=&
    \begin{dcases*}
        \frac{1}{2} &, $0\leq J \leq  \ln{(3/2)}/2$\\ 
        \frac{(\e{2J}+(\e{4J}+4)^{\frac{1}{2}}-\sqrt{2}(\e{2J}(\e{4J}+4)^{\frac{1}{2}}+\e{4J}-6)^{\frac{1}{2}})^{6}}{4096+(\e{2J}+(\e{4J}+4)^{\frac{1}{2}}+\sqrt{2}(\e{2J}(\e{4J}+4)^{\frac{1}{2}}+\e{4J}-6)^{\frac{1}{2}})^{6}} &, $J \geq   \ln{(3/2)}/2$
    \end{dcases*}.
    \label{bare field solution}
\end{eqnarray}
Plugging \eqref{bare
  field solution} into Eq.~\eqref{dphi-BG} and noting that
$\varphi(0)=1/2$ we obtain closed-form expressions for the
interfacial steepness at $x=0$. Similarly, using the definition given by Eq.~\eqref{lch-MF},
we obtain the Cahn-Hilliard interfacial width for the BG
approximation. Results for the interfacial steepness are shown in Fig.~4b in
the  main article with the blue lines and display a strong non-monotonic trend
w.r.t.\ $J$. The broadening of the profile is in sharp contrast to the conclusion drawn by Cahn and Hilliard who write in \cite{scahn1958freeI2}: \emph{``The interface between two coexisting phases is diffuse and its thickness increases with increasing temperature until at the critical temperature (Tc) the
interface is infinite in extent.'' (p266)} Recall that $J$ is expressed in units of $k_{\rm B}T$, and therefore an increase in temperature corresponds to a decrease in $J$. To proof that broadening is a general effect regardless of the lattice we take the strong coupling limit of Eq.~\eqref{dphi-BG}. For $z>2$ and $J\rightarrow\infty$ the nontrivial solutions to Eq.~\eqref{transcendental-eq-BG} are approaching $\chi_{\varphi}\rightarrow 0$ and $\chi_{\varphi}\rightarrow \infty$, resulting in $\varphi_{\rm min}\rightarrow 0$ (as with MF). Plugging this into Eq.~\eqref{dphi-BG} together with $\varphi(0)=1/2$ we obtain
\begin{equation}
    \lim\limits_{J\rightarrow\infty}\varphi^{\prime}(0)=\lim\limits_{J\rightarrow\infty}\pm\sqrt{2(2zJ-z\ln{(\e{2J}+1)}+(z-2)\ln{(2)})/z_{x}\sinh{(2J)}}=0.
\end{equation}
So we find a vanishing interfacial steepness at $x=0$ for any lattice with $z>2$ in the strong coupling limit. For the interfacial width we find
\begin{equation}
    \lim_{J\rightarrow\infty}l_{\rm BG, CH}=\lim_{J\rightarrow\infty}(1-2\varphi_{\rm min})/\varphi^{\prime}(0)=\infty.
\end{equation}
Hence, in the strong-interaction limit the interfacial width diverges for any lattice with $z>2$.
\section{Linear stability analysis}\label{secVII}
\noindent Here we determine the length scales on
which inhomogeneities of the concentration profile are stable, as shown in Sec.~VIC in the main article. We consider a concentration profile of the form $\varphi(\mathbf{x})=\varphi_{0}+a\sin{(\mathbf{q}\cdot\mathbf{x})}$ with $\mathbf{q}=(q_{x},q_{y})^{\rm T}$ and $|a|\ll {\rm min}{(\varphi_{0},1-\varphi_{0})}$. A sinusoidal perturbation is taken to agree with the odd boundary conditions which we imposed for Eq.~\eqref{Euler-Lagrange-MF} and \eqref{Euler-Lagrange-BG}. Expanding the local free energy density and gradient energy coefficient around the homogeneous state up to second order gives 
\begin{equation}
    {\rm f}(\varphi(\mathbf{x}))={\rm f}(\varphi_{0})+a\sin{(\mathbf{q}\cdot\mathbf{x})}{\rm f}^{\prime}(\varphi_{0})+\frac{1}{2}a^{2}\sin^{2}{(\mathbf{q}\cdot\mathbf{x})}{\rm f}^{\prime\prime}(\varphi_{0})+\mathcal{O}(a^{3}),
    \label{expansion-f}
\end{equation}
\begin{equation}
    \frac{1}{2}\nabla \varphi(\mathbf{x})^{\rm T}\boldsymbol{\kappa}(\varphi(\mathbf{x}))\nabla \varphi(\mathbf{x})=\frac{1}{2}a^{2}(\mathbf{q}^{\rm T}\boldsymbol{\kappa}(\varphi_{0})\mathbf{q})\cos^{2}{(\mathbf{q}\cdot\mathbf{x})}+\mathcal{O}(a^{3}),
    \label{expansion-kappa}
\end{equation}
where ${\rm f}^{\prime}(\varphi_{0})\equiv\partial_{\varphi}{\rm f}(\varphi)|_{\varphi_{0}}$ and ${\rm f}^{\prime\prime}(\varphi_{0})\equiv\partial^{2}_{\varphi}{\rm f}(\varphi)|_{\varphi_{0}}$. Now we want to find out when a sinusoidal perturbation decreases the total free energy compared to the uniform concentration profile. Plugging Eqs.~\eqref{expansion-f} and \eqref{expansion-kappa} into Eq.~(27) in the main article and subtracting the free energy density of the uniform concentration gives
\begin{eqnarray}
    F[\varphi(\mathbf{x})]-F[\varphi_{0}]&=&\frac{1}{2L_{x}L_{y}}\int\limits_{-\frac{L_{y}}{2}}^{\frac{L_{y}}{2}}\int\limits_{-\frac{L_{x}}{2}}^{\frac{L_{x}}{2}}[2a\sin{(\mathbf{q}\cdot\mathbf{x})}{\rm f}^{\prime}(\varphi_{0}){+}a^{2}(\sin^{2}{(\mathbf{q}\cdot\mathbf{x})}{\rm f}^{\prime\prime}(\varphi_{0}){+}\cos^{2}{(\mathbf{q}\cdot\mathbf{x})}\mathbf{q}^{\rm T}\boldsymbol{\kappa}(\varphi_{0})\mathbf{q})]dxdy\nonumber\\
    &=&\frac{a^{2}}{4L_{x}L_{y}}\left(L_{x}L_{y}\left({\rm f}^{\prime\prime}(\varphi_{0})+\mathbf{q}^{\rm T}\boldsymbol{\kappa}(\varphi_{0})\mathbf{q}\right){-}(q_{x}q_{y})^{-1}\sin{(q_{x}L_{x})}\sin{(q_{y}L_{y})}\left({\rm f}^{\prime\prime}(\varphi_{0}){-}\mathbf{q}^{\rm T}\boldsymbol{\kappa}(\varphi_{0})\mathbf{q}\right)\right)\nonumber\\
    &=&\frac{a^{2}}{4}\left({\rm f}^{\prime\prime}(\varphi_{0})+\mathbf{q}^{\rm T}\boldsymbol{\kappa}(\varphi_{0})\mathbf{q}\right)+\mathcal{O}\left(\frac{a^{2}}{L_{x}L_{y}}\right),
    \label{dF}
\end{eqnarray}
where in the last line we have taken the large system-size limit
$(L_{x},L_{y})\rightarrow\infty$. To decrease the total free energy
the RHS of Eq.~\eqref{dF} must be negative. Note that $\mathbf{q}^{\rm
  T}\boldsymbol{\kappa}(\varphi_{0})\mathbf{q}\geq0$, and therefore
only ${\rm f}^{\prime\prime}(\varphi_{0})$ can make the RHS
negative.  The region where ${\rm f}^{\prime\prime}(\varphi_{0})<0$ in the $(\varphi_{0},J)$-plane is called the spinodal region, and therefore this process is also known as spinodal decomposition. When ${\rm f}^{\prime\prime}(\varphi_{0})<0$ there is an upper bound on stable wavevectors which is given by 
\begin{equation}
    \mathbf{q}_{\rm crit}^{\rm T}\boldsymbol{\kappa}(\varphi_{0})\mathbf{q}_{\rm crit}=-{\rm f}^{\prime\prime}(\varphi_{0})
    \label{qcrit}
\end{equation}
For a one-dimensional perturbation with $q_{y}=0$ this translates to $q_{\rm crit}=\sqrt{-{\rm f}^{\prime\prime}(\varphi_{0})/\kappa_{x}(\varphi_{0})}$. The critical wavelength given by $\lambda_{\rm crit}=2\pi/q_{\rm crit}$ provides a lower bound on stable wavelengths. We will now determine the properties of $q_{\rm crit}$ and $\lambda_{\rm crit}$ for the MF and BG approximation.
\subsection{Results within Mean Field theory}
\noindent Taking the MF local free energy density and square gradient coefficient defined in Eqs. \eqref{f_MF_per_site}-\eqref{f_uniform_MF_per_site} and plugging them into $q_{\rm crit}$ gives the following result
\begin{equation}
    q_{\rm crit, MF}=\sqrt{-\frac{{\rm f}^{\prime\prime}_{\rm MF}(\varphi_{0})}{\kappa_{{\rm MF},x}}}=\sqrt{\frac{4(z_{x}J_{x}+z_{y}J_{y})-1/(\varphi_{0}(1-\varphi_{0}))}{z_{x}J_{x}}}.
    \label{qcrit-mf}
\end{equation}
For isotropic interaction strength $J_{x}=J_{y}=J$ and inside the spinodal region $zJ\geq1/(4\varphi_{0}(1-\varphi_{0})$ the MF critical wavevector is monotonically increasing with $J$ and for $0<\varphi_{0}<1$ converges to
\begin{equation}
    \lim\limits_{J\rightarrow\infty}q_{\rm crit, MF}=2\sqrt{z/z_{x}}.
\end{equation}
In Fig.~\ref{fig1SM}b we plot Eq.~\eqref{qcrit-mf} for a square lattice
with isotropic interaction strength. The critical wavelength
$\lambda_{\rm crit, MF}$ decreases monotonically with $J$ and
converges within the aforementioned range to the value
$\lim_{J\rightarrow\infty}\lambda_{\rm crit,
  MF}{=}\pi\sqrt{z_{x}/z}$. In Fig.~4f in the main article we
show the MF critical wavelength for a square lattice
with the red line.
\subsection{Results within  Bethe-Guggenheim theory}
\noindent The BG local free energy density and square gradient coefficient are given by Eqs.~(28)-(29) in the main article. For convenience we immediately take the isotropic interaction strength $J_{x}=J_{y}=J$. Plugging the results for the second derivative of the local free energy density (see Eq.~(B17) with $h=0$ in \cite{sblom2022criticality}) into $q_{\rm crit}$ gives
\begin{equation}
    q_{\rm crit}=\sqrt{-\frac{{\rm f}^{\prime\prime}(\varphi_{0})}{\kappa_{x}(\varphi_{0})}}=\sqrt{\frac{2(z-2)(1+4(\e{4J}-1)\varphi_{0}(1-\varphi_{0}))^{\frac{1}{2}}-2z}{z_{x}\varphi_{0}(1-\varphi_{0})(\e{4J}-1)}}.
    \label{qc-BG2}
\end{equation}
Inside the spinodal region $J\geq\ln{((z-1-\varphi_{0}(z-2))(1+\varphi_{0}(z-2))/((z-2)^{2}\varphi_{0}(1-\varphi_{0})))}/4$ (see Eq.~(16) with $h=0$ in \cite{sblom2022criticality}) the BG critical wavevector has a non-monotonic trend and for $0<\varphi_{0}<1$ converges to the value
\begin{equation}
    \lim\limits_{J\rightarrow\infty}q_{\rm crit}=0.
\end{equation}
In Fig.~\ref{fig1SM}c we plot Eq.~\eqref{qc-BG2} for a square lattice with isotropic interaction strength. Similarly the critical wavelength diverges, i.e.\ $\lim_{J\rightarrow\infty}\lambda_{\rm crit}=\infty$. Hence for $0<\varphi_{0}<1$ there exist no finite stable wavelength perturbations in the strong interaction limit. The coupling strength $J^{\dagger}(\varphi_{0})$ where $q_{\rm crit}$ is maximal -- and therefore $\lambda_{\rm crit}$ minimal -- is given by Eq.~(37) in the main article. Remarkably, the maximum of $q_{\rm crit}$ -- and therefore the minimum of $\lambda_{\rm crit}$ -- is independent of the uniform concentration value $\varphi_{0}$ and reads upon plugging Eq.~(37) into Eq.~\eqref{qc-BG2}
\begin{equation}
    q^{\rm max}_{\rm crit}=\frac{2|z-2|}{\sqrt{z_{x}}}\sqrt{\frac{(z(1+\sqrt{z-1}+z/4)-1)^{\frac{1}{2}}-z/2}{z(2+\sqrt{z-1})-2}}.
\end{equation} 
The minimum wavelength is easily obtained by $\lambda^{\rm min}_{\rm
  crit}=2\pi/q^{\rm max}_{\rm crit}$. In Fig.~4f in the
main article we depict the BG critical wavelength for a square lattice with the blue line. The coupling values where
$\lambda_{\rm crit}$ attains a minimum is indicated with the blue
arrow.
\section{Error analysis of the approximate partition functions in finite systems}\label{secVIII}
\begin{figure}
    \centering
    \includegraphics[width=0.75\textwidth]{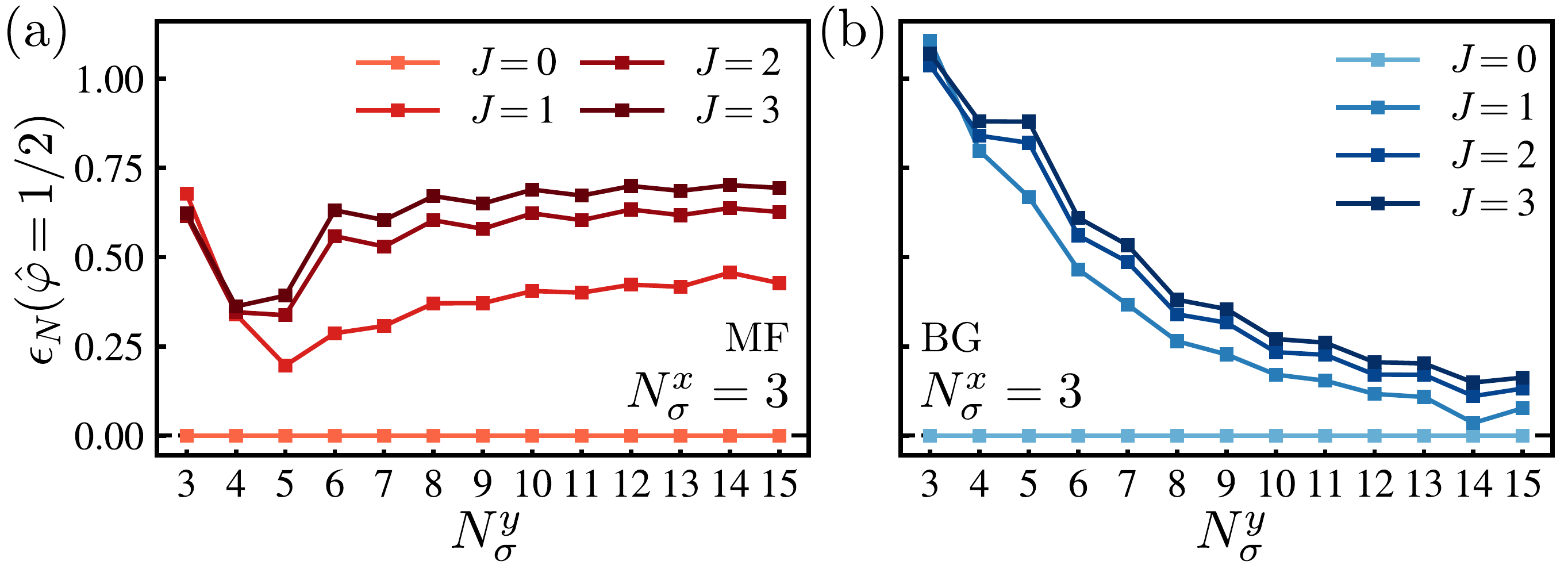}
    \caption{Relative error between the exact and approximated
      partition function obtained with the (a) MF and (b) BG
      approximation for increasing number of spins and various values
      of the coupling strength $J=\{0,1,2,3\}$. The relative
      error in Eq.~\eqref{rel-err} is determined for a square lattice
      composed of
      $(N^{x}_{\sigma}=3)\times(N^{y}_{\sigma}=\{3,...,15\})$ spins
      with periodic boundary conditions in the vertical and
      anti-symmetric boundary conditions in the  horizontal direction, respectively. The total fraction of down spins is fixed to $\hat{\varphi}=1/2$.}
    \label{fig2SM}
\end{figure}

\noindent To probe the accuracy of the MF and BG approximations we compare their partition functions with exact results for the partition function of finite systems. We limit our error analysis to a one-dimensional concentration profile, conform with the majority of results discussed in the main article. For a uniform concentration profile an error analysis between the MF and BG approximation is provided in \cite{sblom2022criticality} (see Fig.~11). For a lattice composed of $N^{x}_{\sigma}\times N^{y}_{\sigma}$ spins, let $\boldsymbol{\varphi}=(\varphi_{1},...,\varphi_{N^{x}_{\sigma}})$ be a vector containing the concentration of downs spins in each column of the lattice. The total concentration of down spins in the lattice is given by $\varphi=||\boldsymbol{\varphi}||_{1}/N^{x}_{\sigma}$. The exact partition function for a fixed concentration profile along the columns is denoted with $Z(\boldsymbol{\varphi})$ and can be computed via
\begin{equation}
    Z(\hat{\boldsymbol{\varphi}})=\sum_{\boldsymbol{\sigma}}\e{-\mathcal{H}(\boldsymbol{\sigma})}\prod_{i=1}^{N^{x}_{\sigma}}\mathbbm{1}_{\hat{\varphi}_{i}}\left[\varphi_{i}\right],
    \label{Z-col}
\end{equation}
where we recall that $\boldsymbol{\sigma}$ denotes the matrix containing all spin configurations, $\mathbbm{1}_{x}\left[z\right]$ is the indicator function of $x$, and $\mathcal{H}(\boldsymbol{\sigma})$ is given by Eq.~(13) in the main article.  The relative error between $Z_{\rm BG, MF}(\boldsymbol{\varphi})$ and $Z(\boldsymbol{\varphi})$ for a fixed total concentration of down spins $\hat{\varphi}$ is defined as
\begin{equation}
    \epsilon_{N}(\hat{\varphi})=\left(\sum_{\boldsymbol{\sigma}}Z(\boldsymbol{\varphi})\left(1-\frac{\ln{(Z_{\rm MF, BG}(\boldsymbol{\varphi}))}}{\ln{(Z(\boldsymbol{\varphi}))}}\right)\mathbbm{1}_{\hat{\varphi}}\left[\varphi\right]\right)/\left(\sum_{\boldsymbol{\sigma}}Z(\boldsymbol{\varphi})\mathbbm{1}_{\hat{\varphi}}\left[\varphi\right]\right).
    \label{rel-err}
\end{equation}
Eq.~\eqref{rel-err} is defined such that differences between
$Z(\boldsymbol{\varphi})$ and $Z_{\rm MF, BF}(\boldsymbol{\varphi})$
attain the largest weight for thermodynamically stable
configurations. In Fig.~\ref{fig2SM} we plot the relative error for the
(a) MF and (b) BG approximation for a finite square lattice composed
of $(N^{x}_{\sigma}=3)\times(N^{y}_{\sigma}=\{3,...,15\})$ spins with
anti-symmetric and periodic boundary conditions in the horizontal and vertical
direction, respectively. Upon increasing the number of spins in the
vertical direction we see that the relative error of the BG
approximation decreases towards zero regardless of the coupling
strength, whereas the MF approximation saturates to a nonzero value
(note that the small system size gives rise to a marked even-odd
dependency). For $J=0$ both approximations are exact and
therefore have zero relative error. The improvement of the BG
approximation with increasing $N^{y}_{\sigma}$ is due to the fact that
it is obtained through a variational principle which is applied in the thermodynamic scaling limit. The MF approximation on the other hand becomes worse with increasing $N^{y}_{\sigma}$ due to the approximation for the fraction of defects given by Eq.~\eqref{Zeta_mf}.
\section{Numerical simulations of the radially symmetric Cahn-Hilliard equation}\label{secIX}
\noindent We study nucleation based on radially symmetric concentration profiles~$\varphi(r)$ in two dimensions. Since critical profiles correspond to stationary points of the free energy~$F$ given by Eqs.~\eqref{f_functional_MF} and \eqref{f_functional_BG}, we next determine minimal free energy paths between the homogeneous state and large droplets. We use a measure for the mass concentrated in the nucleus, $N[\varphi] = \int \tanh(w(\varphi-1/2)) \mathrm{d}V$ with $w=10$, as a reaction coordinate and determine the profile $\varphi(r)$ that minimizes $F$ for a given value~$N_0$ of the constraint using a Lagrange multiplier~$\lambda$.
We thus minimize the constrained free energy
\begin{align}
    F_{\lambda}[\varphi, \lambda] = F[\varphi] - \lambda (N[\varphi] - N_0)
\end{align}
by evolving the corresponding partial differential equations
\begin{subequations}
\begin{align}
    \partial_t \varphi &= \Lambda_D \nabla^2\frac{\delta F_{\lambda}}{\delta \varphi}
\\
    \partial_t \lambda &= -\Lambda_L \frac{\delta F_{\lambda}}{\delta \lambda}
    \;,
\end{align}
\end{subequations}
which corresponds to conserved and non-conserved dynamics with
mobilities $\Lambda_D=10^2$ and $\Lambda_L=10^4$,
respectively. Using this procedure, we determine the
profile~$\varphi(r)$ with Neumann boundary conditions that optimizes $F_{\lambda}$ for each
value $N_0$ of the constraint, which yields the minimal free energy
path. 
The profile with the largest free energy $F$ corresponds to the saddle point and thus to the critical nucleus that we sought.
The corresponding profiles $\varphi(r)$ are shown and analyzed in
Fig.~5 in the main article. Here, the nucleation barrier $\Delta E$ is given by the difference of the energy of the critical nucleus to the energy of the homogeneous state.
\end{document}